\documentclass{aa}  

\usepackage{graphicx}
\usepackage{txfonts}
\usepackage{float}
\usepackage{dblfloatfix}
\usepackage{natbib}
\bibpunct{(}{)}{;}{a}{}{,} 
\usepackage{hyperref}
\hypersetup{colorlinks=true, urlcolor=blue, linkcolor=blue, citecolor=blue}
\usepackage{wrapfig}
\usepackage{comment}
\usepackage{svg}
\begin{document}

   \title{Linking planetary embryo formation to planetesimal formation II:\\ The impact of pebble accretion in the terrestrial planet zone}
   \titlerunning{Linking planetary core formation to planetesimal formation}

   \author{Oliver Voelkel
          \inst{1}
          ,
          Rogerio Deienno
          \inst{2}
          ,
         Katherine Kretke
          \inst{2}
           ,
         Hubert Klahr
          \inst{1}
          }
    \authorrunning{O. Voelkel et al.}

   \institute{\inst{1} Max Planck Institute for Astronomy, Heidelberg, Königstuhl 17, 69117 Heidelberg, Germany\\
   \inst{2} Department of Space Studies, Southwest Research Institute,
Boulder, CO 80302\\
              \\
              \email{voelkel@mpia.de}
             }
  \abstract
   {
   The accretion of pebbles on planetary cores has been widely studied in recent years and is found to be a highly effective mechanism for planetary growth. While most studies assume planetary cores as an initial condition in their simulation, the question how, where and when these cores form is often neglected.   
   }
   {
   We study the impact of pebble accretion during the formation phase and subsequent evolution of planetary embryos in the early stages of circumstellar disk evolution. In doing so we aim to quantify the timescales and local dependency of planetary embryo formation, based on the solid evolution of the disk.
   }
   {
   We connect a one dimensional two population model for solid evolution and pebble flux regulated planetesimal formation to the N-body code LIPAD. In our study we focus on the growth of planetesimals with an initial size of 100$\,$km in diameter by planetesimal collisions and pebble accretion for the first 1 million years of a viscously evolving disk. We compare 18 different N-body simulations in which we vary the total planetesimal mass after 1 million years, the surface density profile of the planetesimal disk, the radial pebble flux and the possibility of pebble accretion.
   }
   {
   Pebble accretion leads to the formation of fewer, but substantially more massive embryos. The area of possible embryo formation is weakly influenced by the accretion of pebbles and the innermost embryos tend to form slightly earlier compared to the simulations in which pebble accretion is neglected. 
   }
   {
    Pebble accretion strongly enhances the formation of super earths in the terrestrial planet region, but it does not enhance the formation of embryos at larger distances.
   }

   \maketitle
%
%
\section{Introduction}
\label{Sec:Instroduction}
\subsection{Physical motivation}
\label{Subsec:physical_motivation}
The accretion of solids and eventually gas \citep{pollack1996formation} on planetary cores is widely used as the standard scenario for planet formation. Most studies in the field of planet formation begin with an initial planetary core that grows by either planetesimal- (\cite{ida2004toward}, \cite{mordasini2012extrasolar}, \cite{EmsenhuberPrepA}, \cite{EmsenhuberPrepB}, \cite{voelkel2020popsynth}) or pebble accretion (\cite{bitsch2015growth}, \cite{Ndugu_2017}, \cite{lambrechts2012rapid}). Recent work included the consistent formation \citep{Lenz_2019} and accretion of planetesimals onto planetary embryos into a global model of planet formation \citep{voelkel2020popsynth}. Despite the improvement, the presence of planetary embryos is still treated as an initial assumption. A fully consistent global model for planet formation however would also have to form planetary embryos based on the previous evolution of the system. Studies that form planetary embryos from planetesimals usually neglect the formation of the planetesimals, by assuming an initial distribution in the disk \citep{levison2015growing,walsh2015planetsimals,carter2015compositional,clement2020embryo}. The study that is presented in this paper is an expansion to our companion paper \citep{voelkel2020embI}, in which we investigated the formation of planetary embryos from a dynamically evolving planetesimal disk and derived a one dimensional, parameterized analytic model for planetary embryo formation. The effect of pebble accretion \citep{ormel2010effect, KlahrBodenheimer2006} on the formation of planetary embryo formation is now added to the same framework in this study.
\\
To motivate our work we discuss the following aspects (often either neglected or not accounted in detail by previous works). One aspect generally neglected in the study of pebble accretion is that the pebble flux in a disk is not a constant, but instead evolves due to radial drift and decays over time. Since pebble accretion relies on the active pebble flux, the time and location at which a planetary embryo is introduced into the simulation is therefore imperative for the evolution of said embryo. The accretion of planetesimals on planetary embryos, as well as planetesimal growth by collisions is sensitive to the size of planetesimals, the local planetesimal surface density and the orbital distance to the star. The evolution and growth of a planet thus strongly depends on its environment, but also the cores themselves are assumed to form from the smaller material in the disk. Modeling the formation of planetary embryos therefore requires an understanding of the local solid evolution of a circumstellar disk. To fully understand the local evolution however, one needs to understand the global evolution of the disk as well, since solids can drift through the circumstellar disk from far out regions. Modeling the formation of planetary embryos in the terrestrial planet region in a self consistent disk therefore requires the understanding of the global formation of planetesimals and evolution of the pebble flux during the time of embryo formation. 
\subsection{Previous work}
\label{Subsec:previous_work}
This study is an extension and of our previous work \cite{voelkel2020embI} in which we studied the impact of the planetesimals surface density and disk mass on the formation of embryos. Our previous study found that the formation of planetary embryos from 100$\,$km planetesimals occurs from the inside out and that the orbital separation of initial embryos converges to $\approx 15$R$_{Hill}$. Our finding confirmed the oligarchic growth nature of the embryo formation process (\cite{kokubo1998oligarchic}, \cite{Kobayashi_2011} and \cite{walsh2019planetesimals} to mention just a few). 
\\
One main result from our first study is that the total number of embryos does not simply increase by introducing more mass in the system. The embryos that exist grow larger, thus increasing their mutual orbital separation. Additionally the formation area within 1 Myr increases for higher disk masses, which leads to a similar number of embryos after 1 Myr for our systems. 
The orbital separation leads to a cumulative number of embryos that increases logarithmic with distance. This behavior is not strongly influenced by the planetesimal surface density profile. 
\\
In \cite{voelkel2020embI} we also introduced an analytic model that succeeded in reproducing the total number, spatial distribution and formation time of planetary embryos when given the same one dimensional planetesimal surface density evolution. 
\\
In our companion paper, we find that the innermost embryos form while planetesimals are still forming as well. This instigates that an active pebble flux exists after the formation of the innermost embryos. The outer embryos form after the formation of planetesimals has mostly vanished. While the accretion of pebbles is not considered in our first study, their presence is promising for planetary growth.
\\
In addition to our previous work we now introduce the accretion of pebbles onto planetesimals and planetary embryos. Studies regarding the evolution of a planetary system from planetesimals and pebbles in the LIPAD \citep{levison2012lagrangian} Code have already been conducted by \cite{kretke2014challenges}. In contrast to what has been studied in \cite{kretke2014challenges} we introduce the planetesimal over time based on their formation of a one dimensional planetesimal formation model described in Sect. \ref{Sec:PPE}.
\subsection{The goal of this study}
\label{Subsec:study_goal}
Within this study we connect a global model for the evolution of a circumstellar disk, that involves the formation and drift of pebbles, as well as the pebble flux regulated formation of planetesimals with N-body simulations. The N-body then tracks their subsequent growth and dynamical evolution. Using this framework we study a wide range of parameters to investigate their individual contribution on the formation of planetary embryos and the evolution of planetary systems in the terrestrial planet region. This paper is an addition to our previous study \citep{voelkel2020embI} in which we study the impact of the planetesimal surface density profile and disk mass on the formation of planetary embryos in the terrestrial planet region. Additionally to the formation of planetesimals, we now introduce a radial pebble flux and the possibility of pebble accretion into our framework. The evolution of the pebble flux stems from the same disk evolution that also forms the planetesimals within the N-body simulation. Comparing our results from this study with our previous study, we present 18 different N-body siumulations in which we vary the planetesimal surface density profile, the total mass in planetesimals and the total pebble flux.
\\
In Sect. \ref{Sec:PPE} we summarize the theory behind our approach and explain the numerical setup in Sect. \ref{Sec:Simulation Setup}. Sect. \ref{Sec:Numerical_Results} presents the results that are discussed in Sect. \ref{Sec:Discussion}. Sect. \ref{Sec:Summary} summarizes our findings and gives an outlook to future work. 
\section{Pebbles, planetesimals and embryos}
\label{Sec:PPE}
The goal of this study is to comprehensively model the formation and early dynamical evolution of planetary embryos, following an initial population of dust as it is converted into pebbles and planetesimals. We specifically focus on investigating how the accretion of pebbles impacts this formation process. The framework that we have chosen to make this possible is split up into two parallel sub-processes. We first compute the viscous evolution of a circumstellar gas disk including its solid evolution and planetesimal formation. The qualitative evolution of the solids will serve as a proxy for planetesimal formation and the pebble flux to be included in the N-body simulations. This way the N-body simulation runs with the planetesimals and pebbles that have been formed using the one dimensional approach, while continuing to compute their growth via collision and accretion. 
\\
A detailed description of the pebble flux regulated planetesimal formation model and the two population solid evolution model can be found in \cite{Lenz_2019} and \cite{birnstiel2012simple}. Our approach of coupling the one dimensional planetesimal formation model to the N-body simulation in LIPAD \citep{levison2012lagrangian}, as well as a detailed description of the physical models is described in our previous work \citep{voelkel2020embI}. In the following we give a brief summary of the underlying physical principles.
\subsection{Planetesimal formation and pebble evolution}
\label{Subsec:Pts_formation}
Our framework uses the two population solid evolution approach from \cite{birnstiel2012simple} to compute the dust and pebble evolution of a viscously evolving circumstellar disk \citep{shakura1973black} and the pebble flux regulated planetesimal formation model by \cite{Lenz_2019}. This framework has recently been used to study the impact of planetesimal formation on the formation of planets \citep{voelkel2020popsynth} and was applied in our companion paper \citep{voelkel2020embI}. 
\\
The two population model uses a parameterized mass relation between a small and a large population of solids in the disk, defined by their Stokes number. The small particles ($St \ll 1 $) are coupled to the dynamic motion of the gas and can be seen as dust, while the larger particles ($St \sim 1 $) are detached from the gas motion and can be seen as pebbles. The parameter that separates the two populations has been derived by fitting the two population approach to larger coagulation based simulations of grain growth \citep{Birnstiel_2010}. Planetesimals then form proportional to the radial pebble flux \citep{Lenz_2019}. The planetesimal formation model assumes that particle traps can appear at any location in the disk and last for a given lifetime. The model assumes that a fraction of the radial pebble flux that drifts through a particle trap can be transformed into planetesimals. Planetesimals form with an initial size of 100$\,$km in diameter \citep{Klahr2020, abod2019mass, Johansen2009}, which leads to a threshold mass that has to be reached in order to form planetesimals in this one dimensional approach. The approach itself does not specify what underlying mechanism/instability (e.g. streaming instability, Kelvin Helmholtz instability) drives the formation of planetesimals, it is a model independent framework that forms planetesimals based on the radial pebble flux. 
\subsection{Embryo formation}
\label{Subsec:Embryo_formation}
We define a planetary embryo as an object with at least a lunar mass (M$_{e} = 0.0123$M$_{\oplus}$) in our study. Growing an embryo from 100 km-sized planetesimals (with a bulk density of $\rho = 2g/cm^3$) requires more than 5 orders of magnitude of growth). This would require hundreds of thousands of planetesimals to form a single embryo via collisions, making this problem computationally unfeasible for classical numerical integrators. Thus, in order to tackle this problem we use the code known as LIPAD \citep{levison2012lagrangian}. LIPAD is a lagrangian code that uses the concept of tracer particles to follow the dynamical/collisional/accretional evolution of a huge number of sub-km-sized planetesimals all the away to become planets. We direct the reader to \cite{voelkel2020embI} for a detailed description on how we convert the 1-D solid evolution outcomes into tracers, as well as \cite{levison2012lagrangian}; \cite{kretke2014challenges}; \cite{walsh2016terrestrial}; \cite{walsh2019planetesimals}; \cite{deienno2019energy}: \cite{deienno2020collisional} for a series of previous applications of LIPAD. 
\\
Our study introduces planetesimal and pebble tracer particles and computes their growth by planetesimal collisions and pebble accretion. Tracer particles are represented by three quantities: mass, physical radius and bulk density. These three quantities relate to each other as $n_{pl} = m_{tr} / [ (4./3.) \rho r_{pl}^3 ]$. Here, $n_{pl}$ is the number of planetesimals represented by a single tracer particle, $m_{tr}$ is the tracer constant mass, $\rho$ its constant bulk density and $r_{pl}$ the planetesimal size, that the tracer will represent. This implies that the number of planetesimals represented by a single tracer is larger for smaller planetesimals. It also implies that as planetesimals growth due to their collisional evolution/accretion, they are less represented by a single tracer. As a result, once a planetesimal grows to the point where a tracer will represent only one object (a Moon sized object in our case), this tracer is promoted to an embryo and is then treated as an individual N-body object in the simulation. The promotion of a planetesimal tracer particle to a planetary embryo in LIPAD is what we define the initial formation of a planetary embryo.
\subsection{Pebble accretion}
\label{Subsec:Pebble_accretion}
The fundamental difference to part I of our study lies in the accretion of pebbles onto planetesimals and planetary embryos. In the following we will briefly explain the concept of pebble accretion based on \cite{ormel2010effect} and \cite{lambrechts2012rapid}. A detailed description on how pebble accretion is implemented in LIPAD can be found in \cite{kretke2014challenges}. When we refer to pebble accretion, we talk about the accretion of particles on bodies that is strongly enhanced by gas drag. For this to occur, several conditions need to be met. The stopping timescale of the particle that is to be accreted must be long compared to the timescale of deflection by the target's object gravity. More specifically, the gravitational encounter timescale must be shorter than four times the stopping time
\begin{align}
    v_{rel} \frac{b^2}{G M_p} < 4 t_s
\end{align}
with $G$ as the gravitational constant and $t_s$ the stopping time.
$v_{rel}$ is given as the relative velocity of the particle and the planetesimal/planetary embryo of mass $M_p$. The impact parameter $b^2$ can then be expressed as 
\begin{align}
    b < \Tilde{R}_C = \left(    \frac{4 G M_p t_s} {v_s}        \right)^{1/2} .
\end{align}
The second criterion states that the stopping time of the particle must be short compared to the time it takes for the particle to drift past the target. The impact parameter for when a particle is deflected by $90^{\circ}$ then gives
\begin{align}
    b = b_{90^{\circ}} = \frac{G M_p}{v_{rel}^2} .
\end{align}
In summary, the first criterion states that small dust cannot contribute to pebble accretion because it is too strongly coupled to the motion of the gas, while the second criterion illustrates why larger objects like planetesimals do not benefit from gas drag. The critical crossing time scale can then be defined as 
\begin{align}
    t_{s,*} = \frac{b_{90^{\circ}}}{v_{rel}} = \frac{G M_p}{v_{rel}^3} . 
\end{align}
In the LIPAD simulation, pebbles radially drift inwards. The decision whether a pebble can be accreted by an object is made if the particle is within the Hill radius of the object and under the condition that 
\begin{align}
    b < R_C = \Tilde{R}_C \exp{\left[ - \left( \frac{t_s}{4 t_{s,*}} \right)^{\gamma}  \right]}
\end{align}
with $\gamma = 0.65$.
Pebbles enter the N-body simulation in the form of pebble tracers \citep{kretke2014challenges}.
\section{Simulation Setup}
\label{Sec:Simulation Setup}
\begin{figure}
    \centering
    \includegraphics[width=1.0\linewidth]{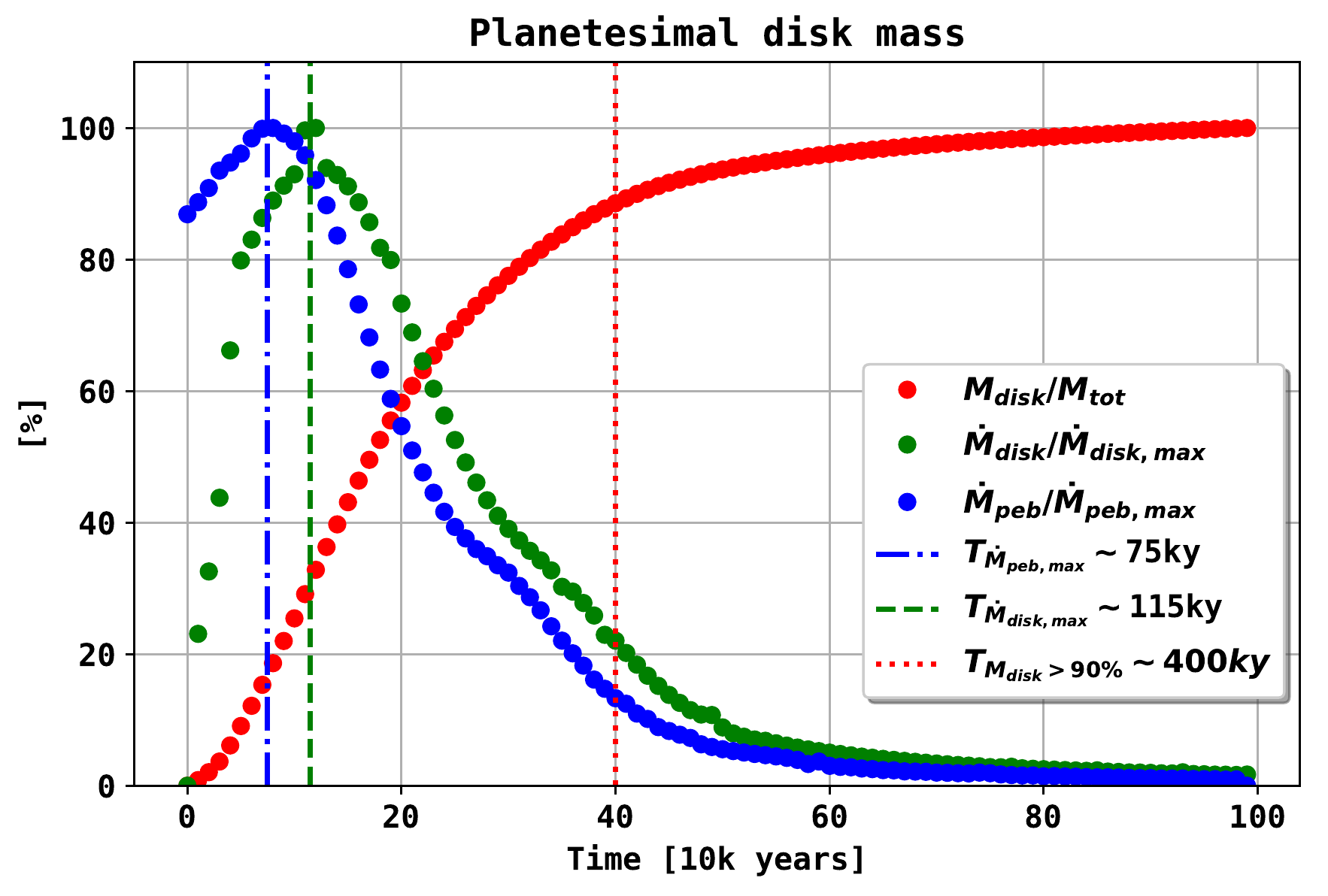}
    \caption{Percentage change of the planetesimal disk mass $\dot{M}_{disk}$ , the total disk mass $M_{disk}$ and the radial pebble flux at 5$\,$au. The disk mass (red dots) is normalized by the total disk mass after $10^6$ years. The green dots indicate the disk mass increase every $10^4$ years ($\dot{M}_{disk}$), normalized by the maximum mass change ($\dot{M}_{disk, max}$). The blue dots indicate the pebble flux every $10^4$ years ($\dot{M}_{peb}$), normalized by the maximum pebble flux ($\dot{M}_{peb, max}$). 
    We find that $\sim 90 \%$ of planetesimals have formed within 400$\,$ky with a peak in the pebble flux at $\sim 75$ky and another in planetesimal formation at $\sim 115$ky.
    }
    \label{Fig:formation_rate}
\end{figure}

The setup of our present study is an expansion of our previous work \citep{voelkel2020embI} and is described there in greater detail, but for the purposes of this work we briefly describe the model setup here. We compute the first 1 Myr of a viscously evolving disk including the two population solid evolution and pebble flux regulated planetesimal formation model from Sect. \ref{Sec:PPE}. The mass rate of planetesmimal formation is then given as an input to the LIPAD N-body simulation in terms of a corresponding number of planetesimal tracers every 10$\,$kyr. With our setup we study the evolution of planetary embryo formation for 18 different systems in which we vary the total planetesimal disk mass after 1 Myr, the planetesimal surface density profile and the total pebble flux. The total planetesimal masses after 1 Myrs are given by 6$\,$M$_{\oplus}$, 13$\,$M$_{\oplus}$ and 27$\,$M$_{\oplus}$. The planetesimal surface density profile varies by $\Sigma_P \propto r^{-1.0}$, $\Sigma_P \propto r^{-1.5}$ and $\Sigma_P \propto r^{-2.0}$. Our study individually compares systems in which pebble accretion is active to those in which it is ignored. In addition to our previously published work \citep{voelkel2020embI} we introduce a radial pebble flux into the LIPAD simulation. Pebbles are placed outside the outer edge of our computational domain at 5$\,$au. The total mass of the pebble flux over 1 Myr is varied by 57.7$\,$M$_{\oplus}$ in the 6$\,$M$_{\oplus}$ case, 115.8$\,$M$_{\oplus}$ in the 13$\,$M$_{\oplus}$ case and 232.5$\,$M$_{\oplus}$ in the 27$\,$M$_{\oplus}$ case. The corresponding mass is introduced over 1 Myr into the simulation in the form of pebble tracers. These tracers do not contribute to the planetesimal formation, but can be accreted by the planetesimal tracers and embryos. The qualitative evolution of the pebble flux at 5$\,$au is taken from our one dimensional solid evolution model as well, similar to the formation of the planetesimal disk. 
The change of the disk mass, the disk formation rate and the radial pebble flux at 5$\,$au that is used in our setups is shown in Fig. \ref{Fig:formation_rate}.
\section{Numerical results}
\label{Sec:Numerical_Results}
 In our analysis we focus on the time and semimajor axis evolution (Fig. \ref{Fig:Emb_form_LIPAD_6_ME}. - Fig. \ref{Fig:Emb_form_LIPAD_27_ME}), the individual embryo masses (Fig. \ref{Fig:Embryo_masses}), the total number and mass in embryos over time (Fig. \ref{Fig:Embryo_number}), the mean orbital separation of embryos over time (Fig. \ref{Fig:Orbital_separation}) and the cumulative distribution of embryos in the disk (Fig. \ref{Fig:Cumulative_number}).
\subsection{Embryo formation}
\label{Subsec:Embryo_formation}
\begin{figure*}[]
\centering
\includegraphics[width=1.0\linewidth]{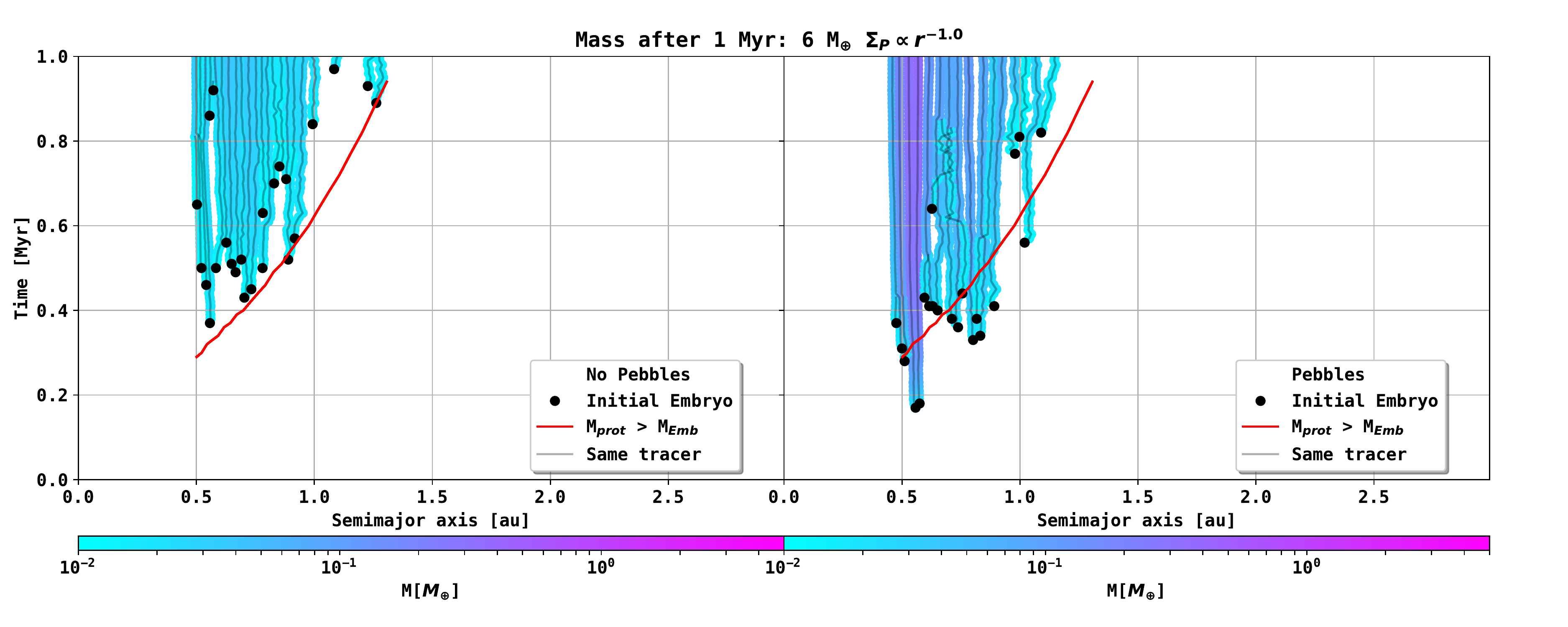} \\
\includegraphics[width=1.0\linewidth]{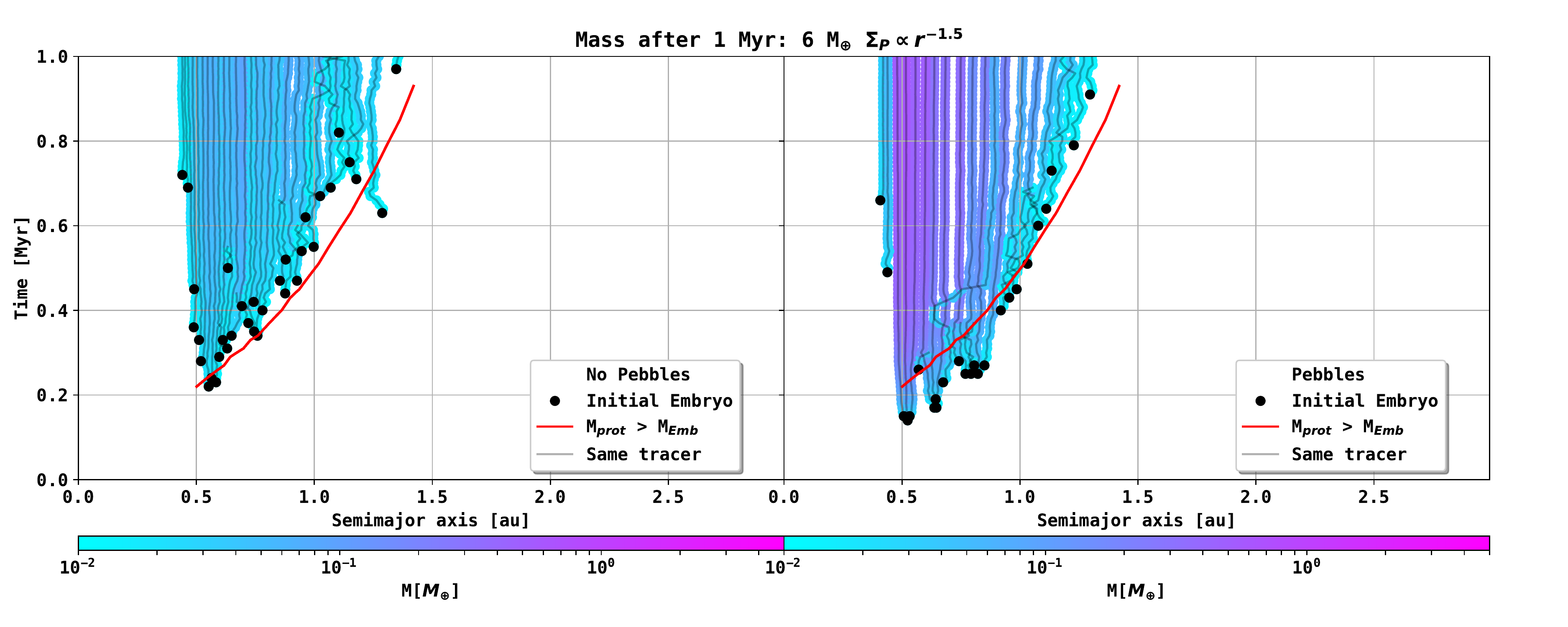} \\
\includegraphics[width=1.0\linewidth]{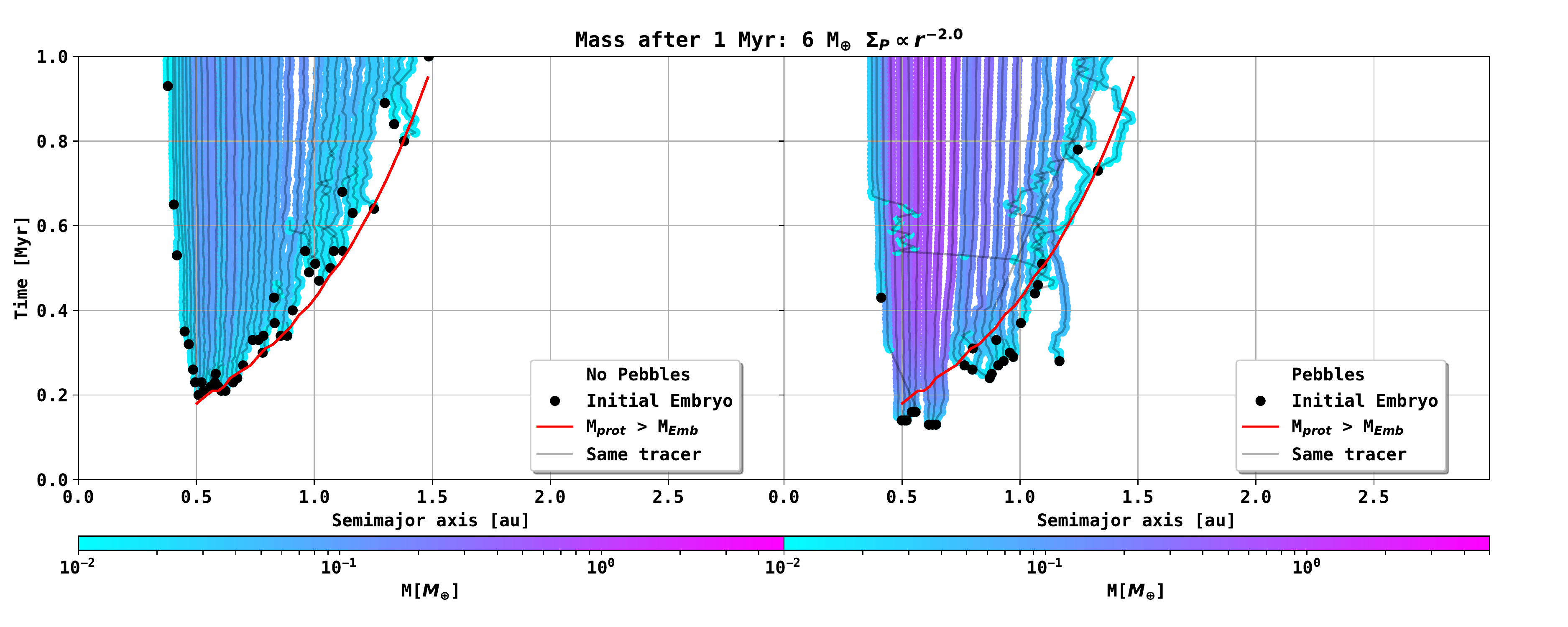} \\
\caption{\small  Time over semimajor axis evolution of the N-body simulation in LIPAD. The time and location at which an object has first reached lunar mass is indicated by the black dots in the plot. The subsequent growth of the embryo is tracked and connected with the grey lines, its mass is given by the colorbar. The mass in planetesimals after 1 Myr is given by 6 M$_{\oplus}$ in these runs, the planetesimal surface density slope is varied ($\Sigma_P \propto r^{-1.0}$, $\Sigma_P \propto r^{-1.5}$ , $\Sigma_P \propto r^{-2.0}$ ). The left panels show the system without pebble accretion. The right panels show the system in which pebble accretion is included. The red line indicates the time after which the analytic model presented in \cite{voelkel2020embI} states that embryo formation is possible.
}
\label{Fig:Emb_form_LIPAD_6_ME}
\end{figure*}

\begin{figure*}[]
\centering
\includegraphics[width=1.0\linewidth]{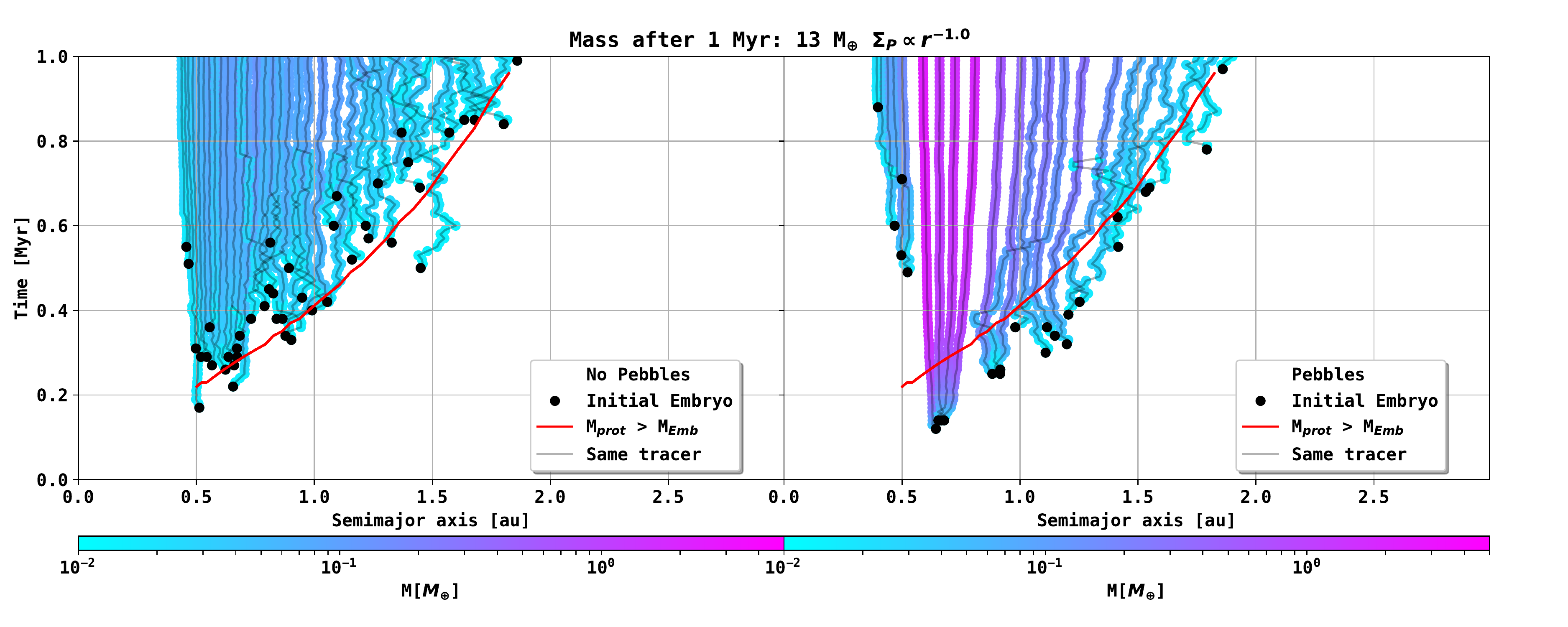} \\
\includegraphics[width=1.0\linewidth]{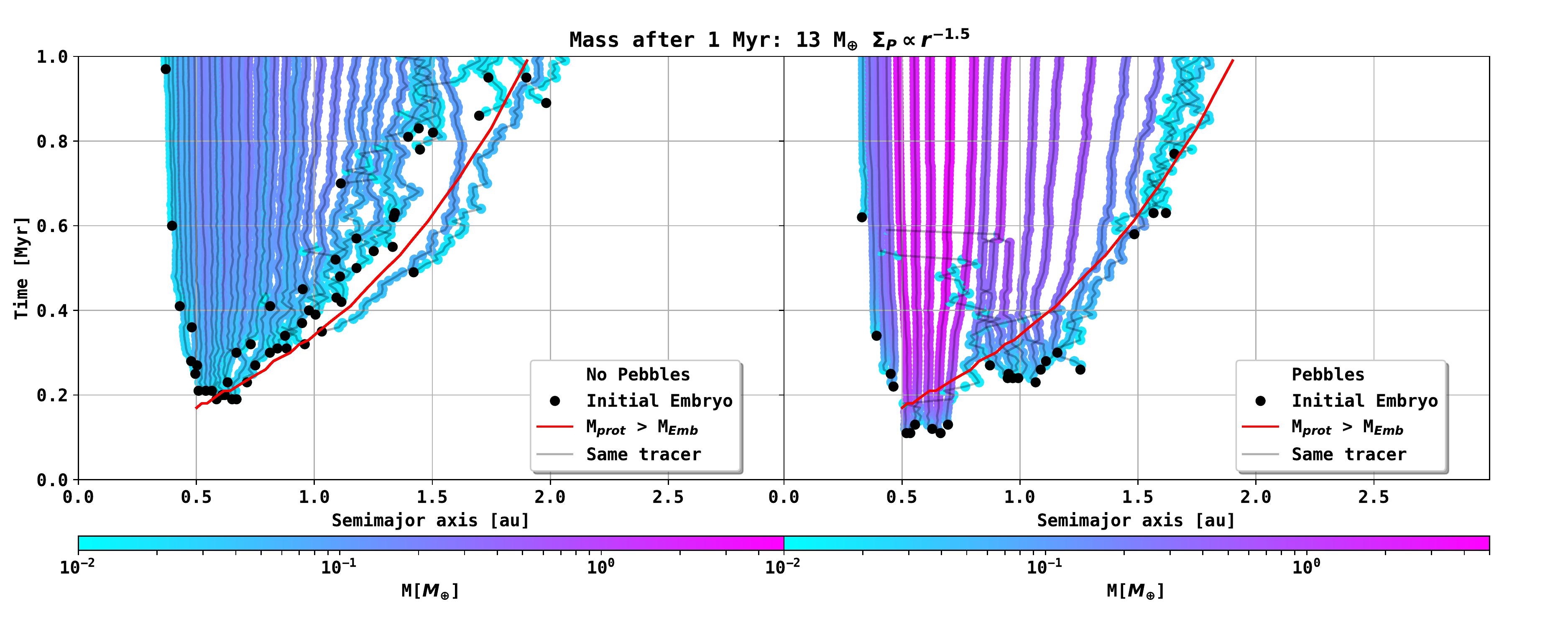} \\
\includegraphics[width=1.0\linewidth]{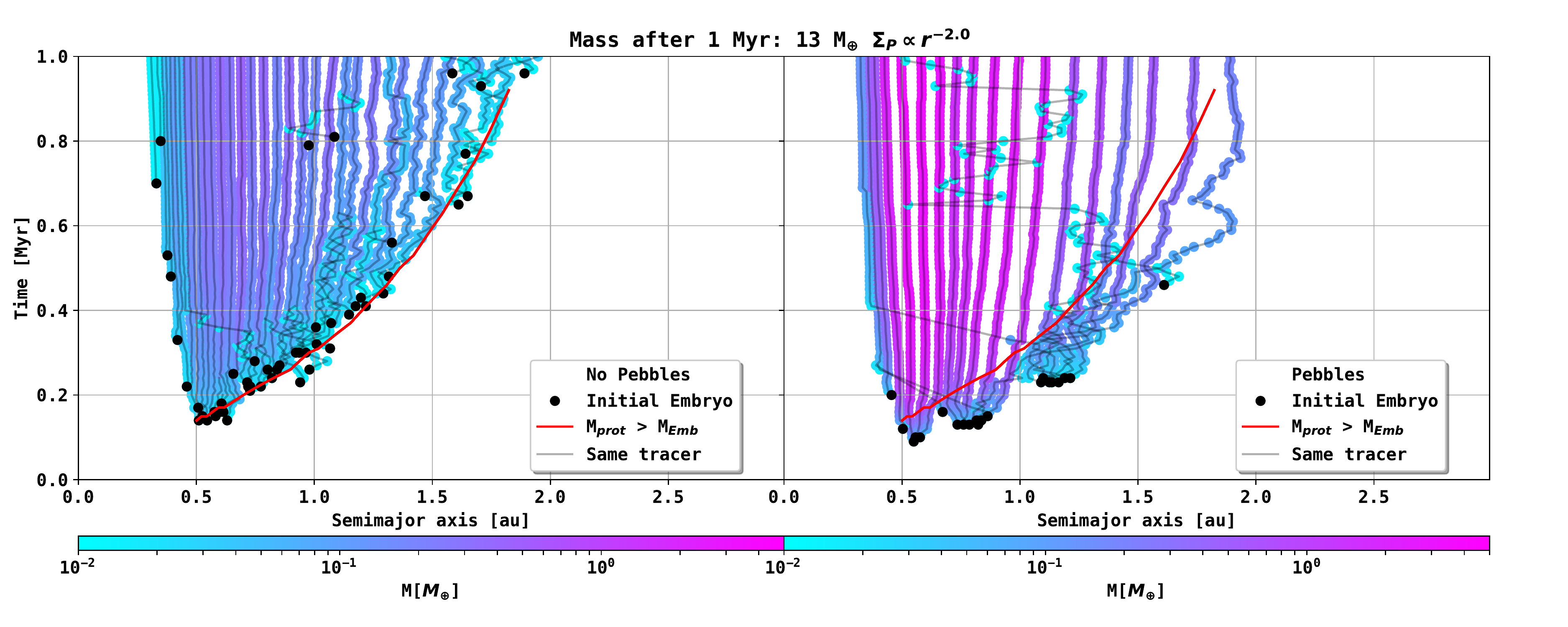} \\
\caption{\small  Time over semimajor axis evolution of the N-body simulation in LIPAD. The time and location at which an object has first reached lunar mass is indicated by the black dots in the plot. The subsequent growth of the embryo is tracked and connected with the grey lines, its mass is given by the colorbar. The mass in planetesimals after 1 Myr is given by 13 M$_{\oplus}$ in these runs, the planetesimal surface density slope is varied ($\Sigma_P \propto r^{-1.0}$, $\Sigma_P \propto r^{-1.5}$ , $\Sigma_P \propto r^{-2.0}$ ). The left panels show the system without pebble accretion. The right panels show the system in which pebble accretion is included. The red line indicates the time after which the analytic model presented in \cite{voelkel2020embI} states that embryo formation is possible.
}
\label{Fig:Emb_form_LIPAD_13_ME}
\end{figure*}

\begin{figure*}[]
\centering
\includegraphics[width=1.0\linewidth]{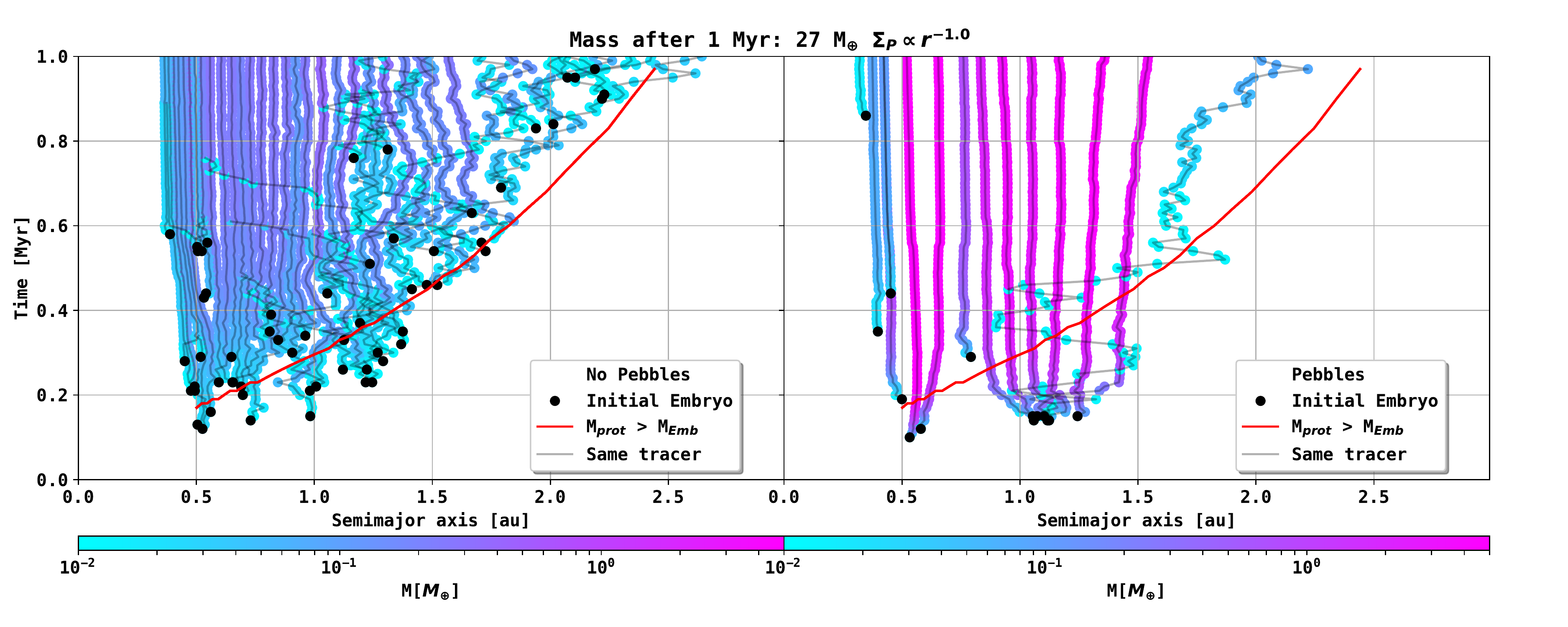} \\
\includegraphics[width=1.0\linewidth]{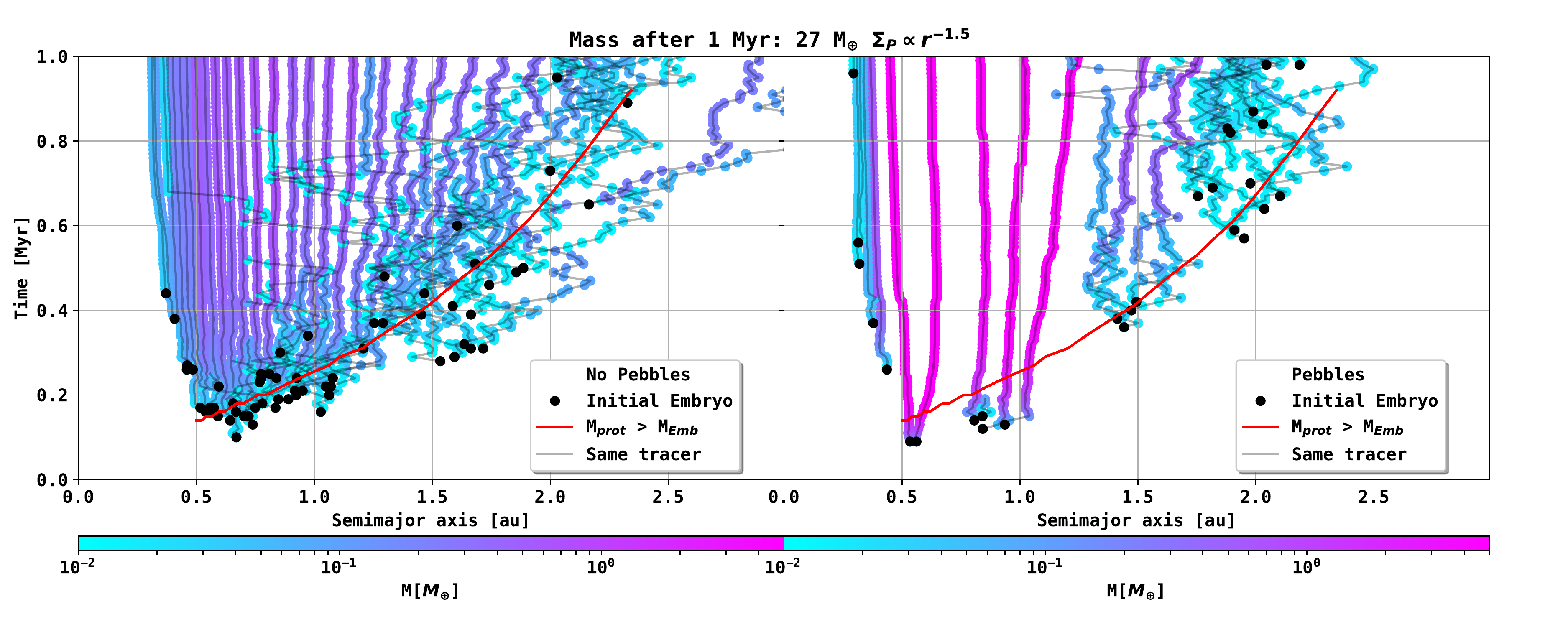} \\
\includegraphics[width=1.0\linewidth]{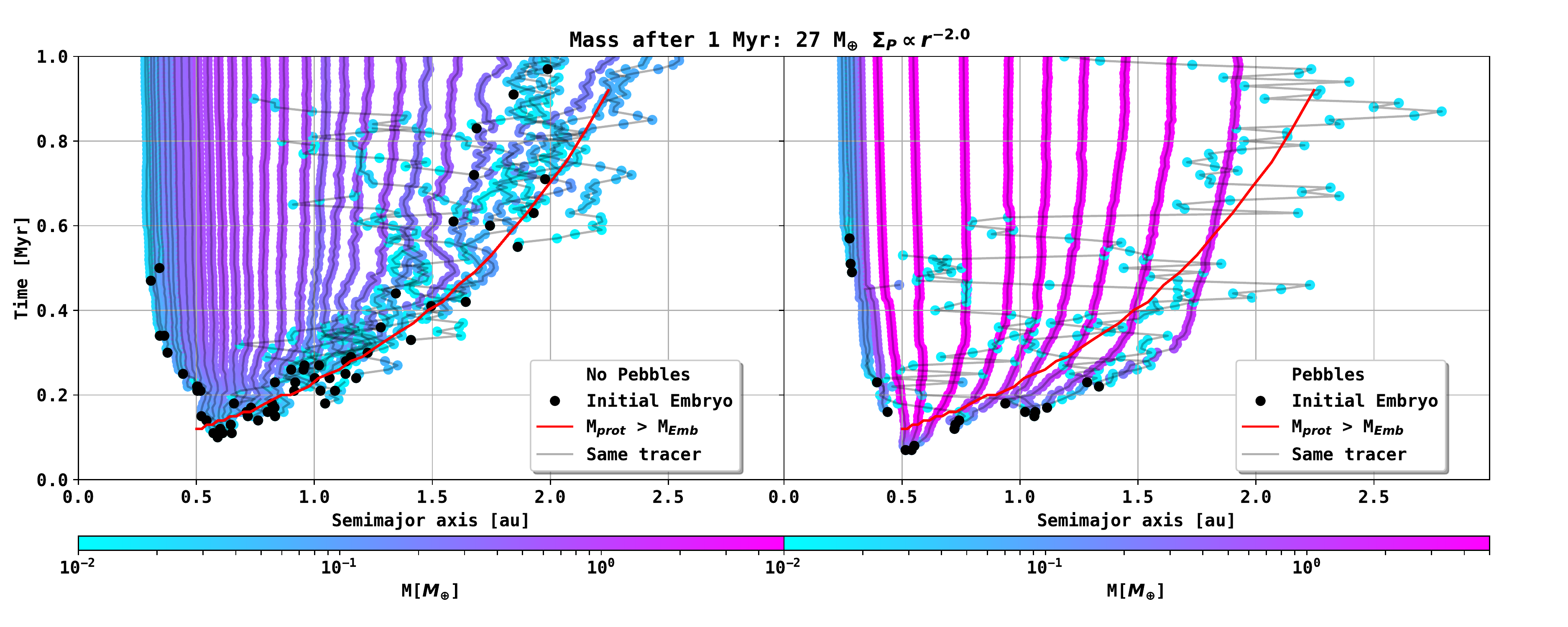} \\
\caption{\small  Time over semimajor axis evolution of the N-body simulation in LIPAD. The time and location at which an object has first reached lunar mass is indicated by the black dots in the plot. The subsequent growth of the embryo is tracked and connected with the grey lines, its mass is given by the colorbar. The mass in planetesimals after 1 Myr is given by 27 M$_{\oplus}$ in these runs, the planetesimal surface density slope is varied ($\Sigma_P \propto r^{-1.0}$, $\Sigma_P \propto r^{-1.5}$ , $\Sigma_P \propto r^{-2.0}$ ). The left panels show the system without pebble accretion. The right panels show the system in which pebble accretion is included. The red line indicates the time after which the analytic model presented in \cite{voelkel2020embI} states that embryo formation is possible.
}
\label{Fig:Emb_form_LIPAD_27_ME}
\end{figure*}

 Fig. \ref{Fig:Emb_form_LIPAD_6_ME} - Fig. \ref{Fig:Emb_form_LIPAD_27_ME} show the time, mass and semimajor axis evolution of planetary embryo formation with the LIPAD code. The toal mass after 1 Myr in planetesimals is given as $6 M_{\oplus}$ (Fig. \ref{Fig:Emb_form_LIPAD_6_ME}), $13 M_{\oplus}$ (Fig. \ref{Fig:Emb_form_LIPAD_13_ME}) and $27 M_{\oplus}$ (Fig. \ref{Fig:Emb_form_LIPAD_27_ME}). The simulations in which pebble accretion is not included (left panels) were taken from our previous work \citep{voelkel2020embI} and serve as comparison in this study. The panels on the right always show the same system in which pebble accretion is included. The color map shows the mass of the objects that are considered embryos in the LIPAD simulations, while the black dots refer to the location and time at which a tracer particle has been promoted to a planetary embryo. The black dots can therefore be interpreted as the initial formation of embryos. In addition to this we define the term  'active' embryos. This term refers to all objects above embryo mass at a given time. Every active embryo used to be an initial embryo, however not every initial embryo remains in the system due to mergers. The individual embryos are connected by a grey line for clarity.
\subsection{Embryo masses}
\label{Subsec:Embryo_masses}
\begin{figure*}[]
\centering
\begin{minipage}{.33\textwidth}
  \centering
  \includegraphics[width=1.0\linewidth]{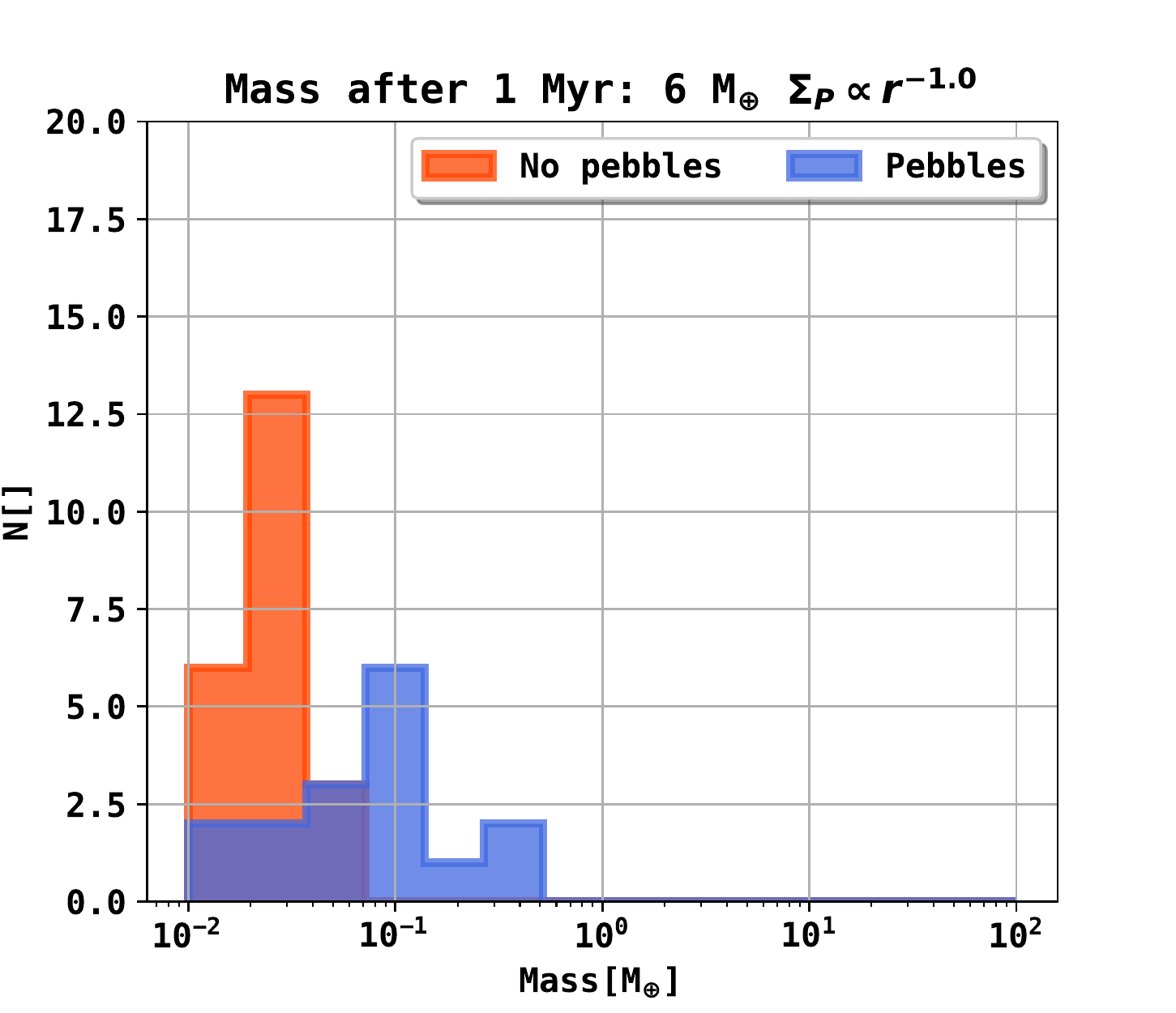}
\end{minipage}
\begin{minipage}{.33\textwidth}
  \includegraphics[width=1.0\linewidth]{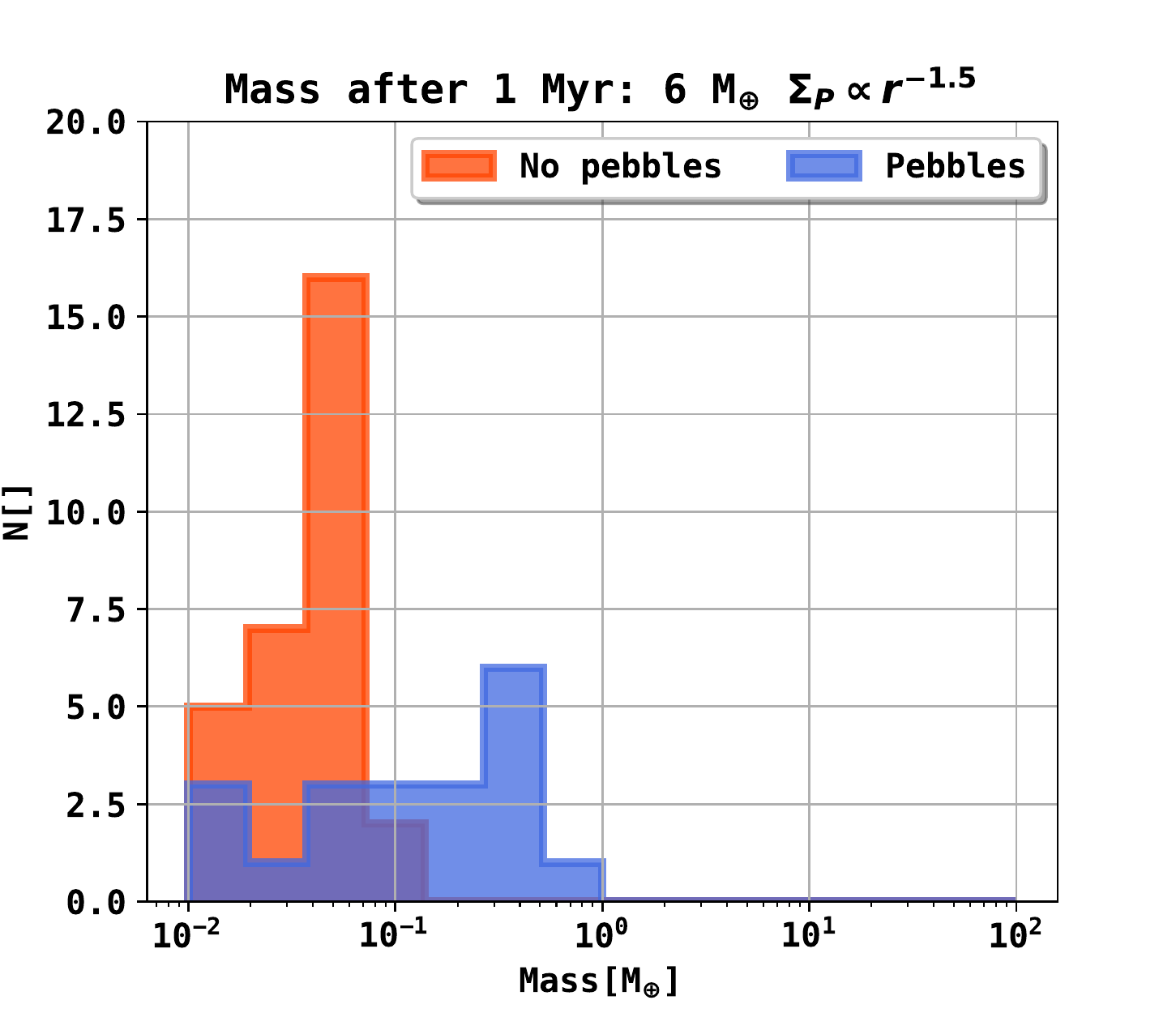}
\end{minipage}%
\begin{minipage}{.33\textwidth}
  \includegraphics[width=1.0\linewidth]{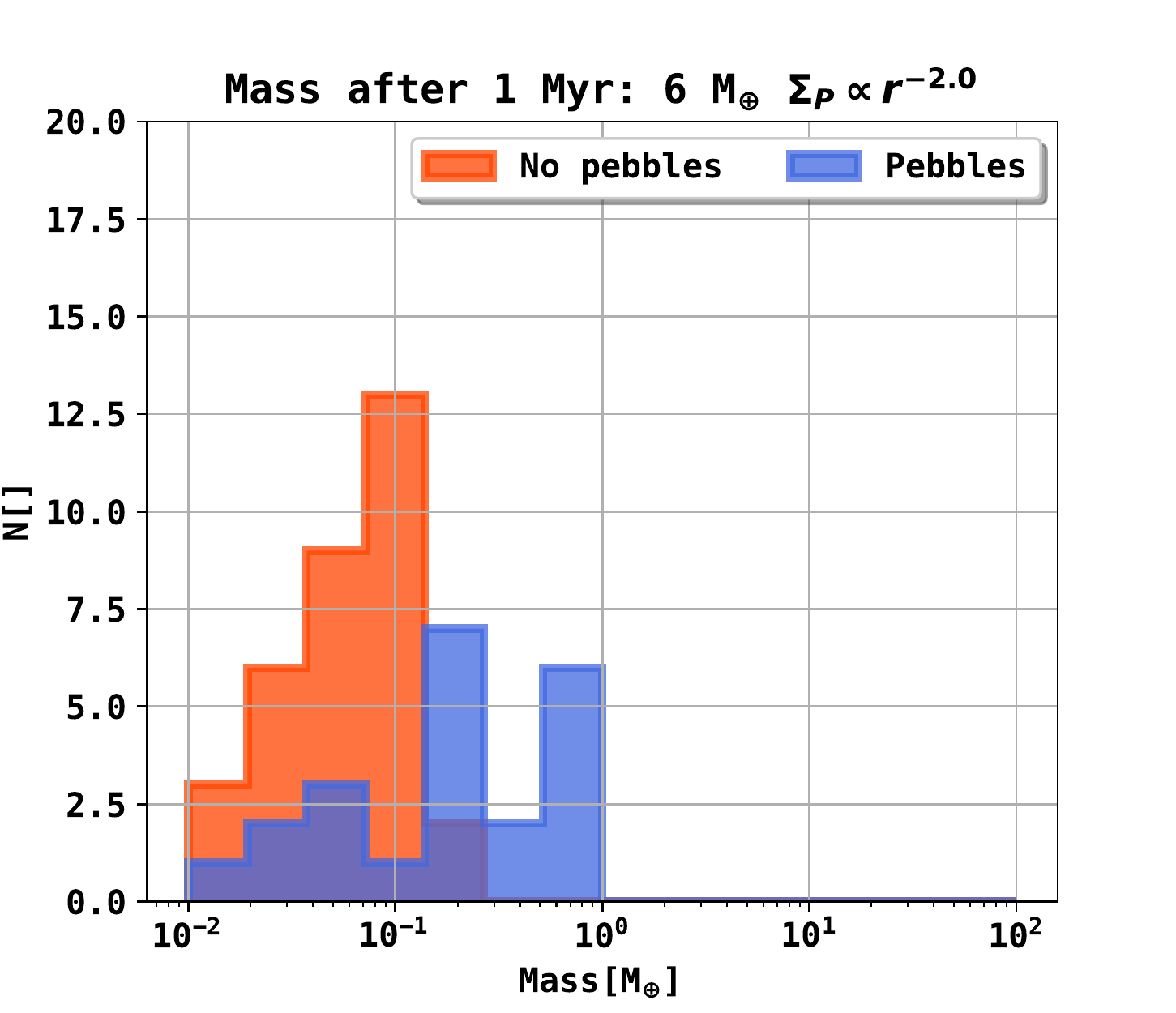}
\end{minipage}%
\\
\begin{minipage}{.33\textwidth}
  \centering
  \includegraphics[width=1.0\linewidth]{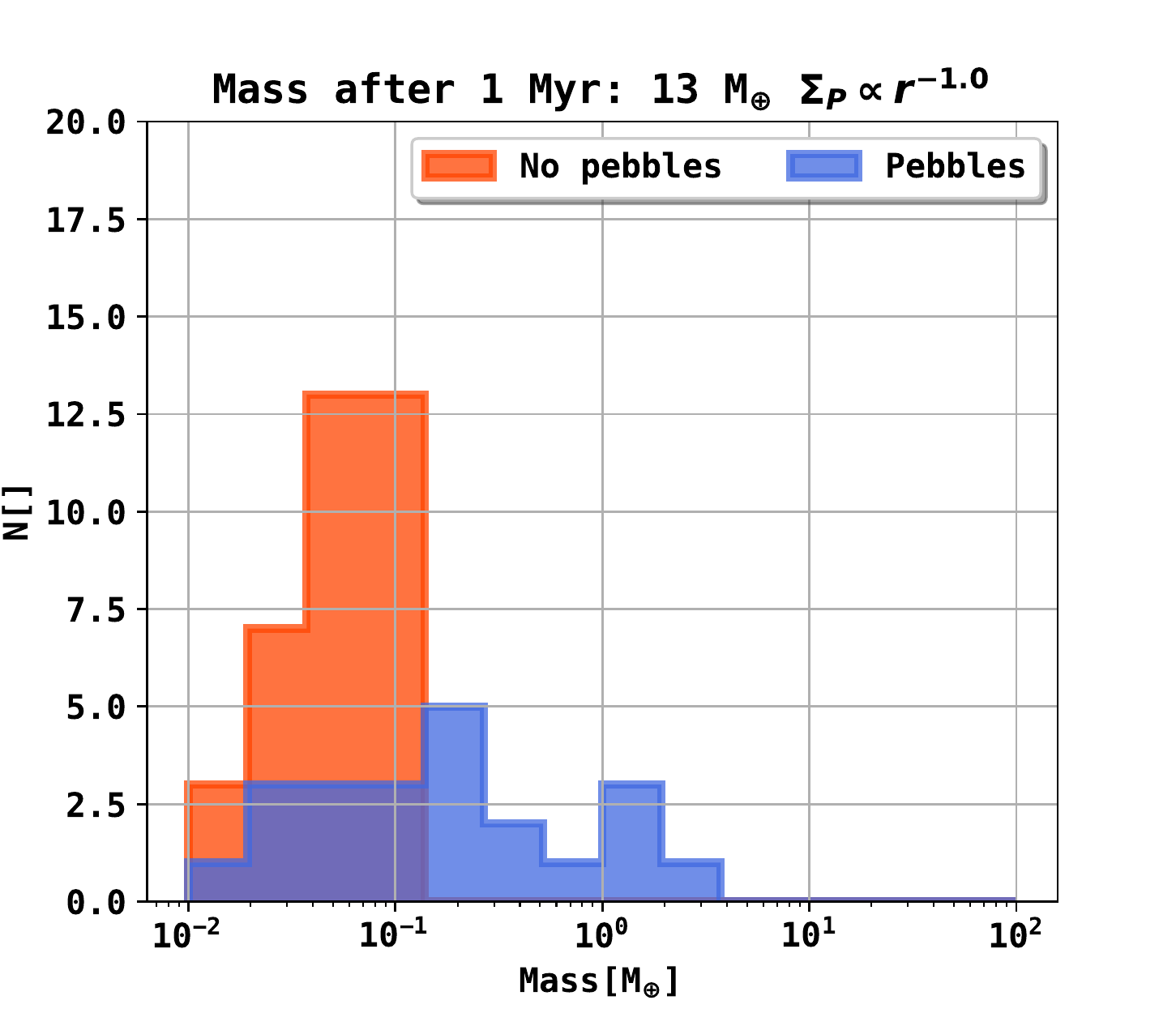}
\end{minipage}
\begin{minipage}{.33\textwidth}
  \includegraphics[width=1.0\linewidth]{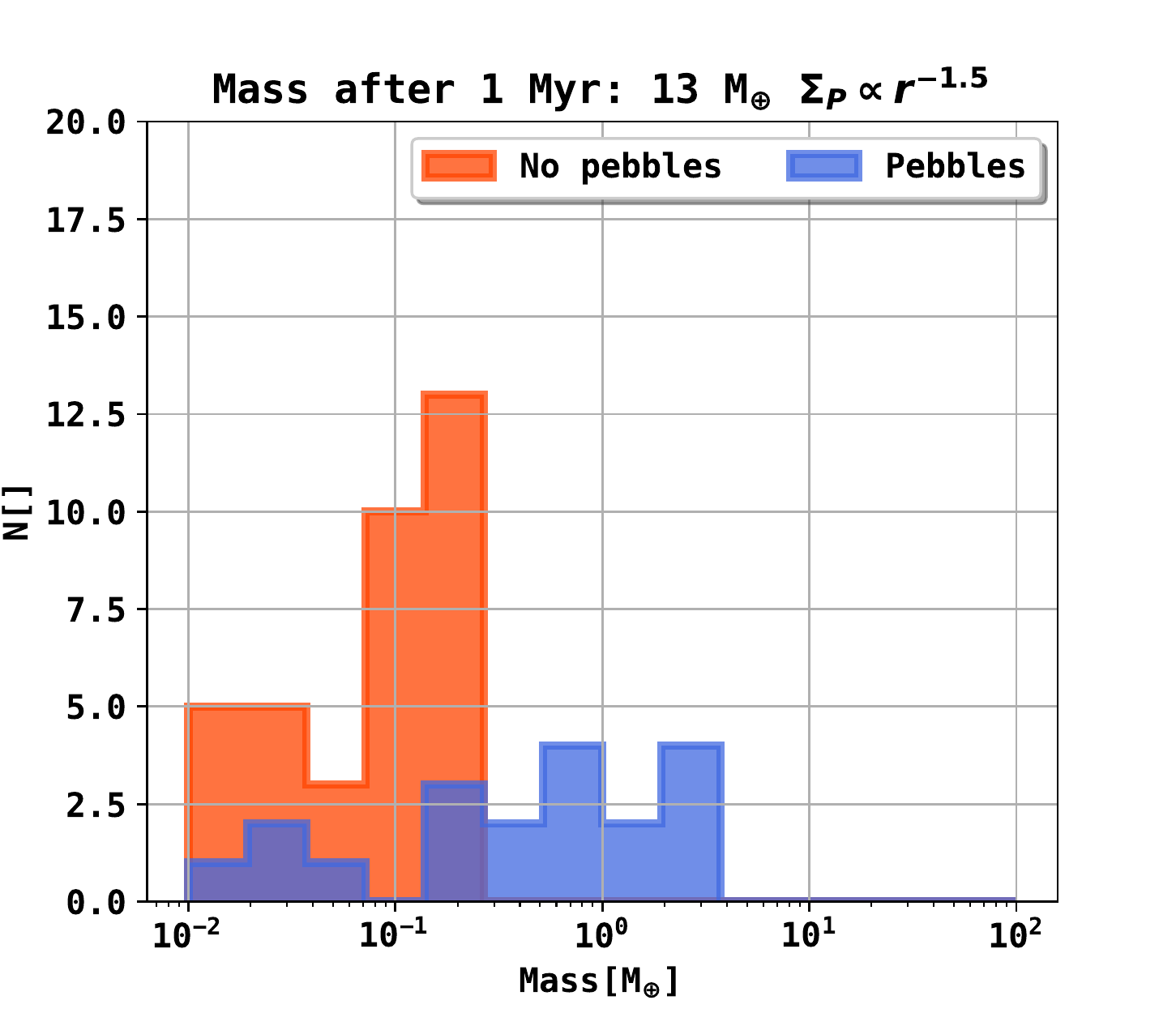}
\end{minipage}%
\begin{minipage}{.33\textwidth}
  \includegraphics[width=1.0\linewidth]{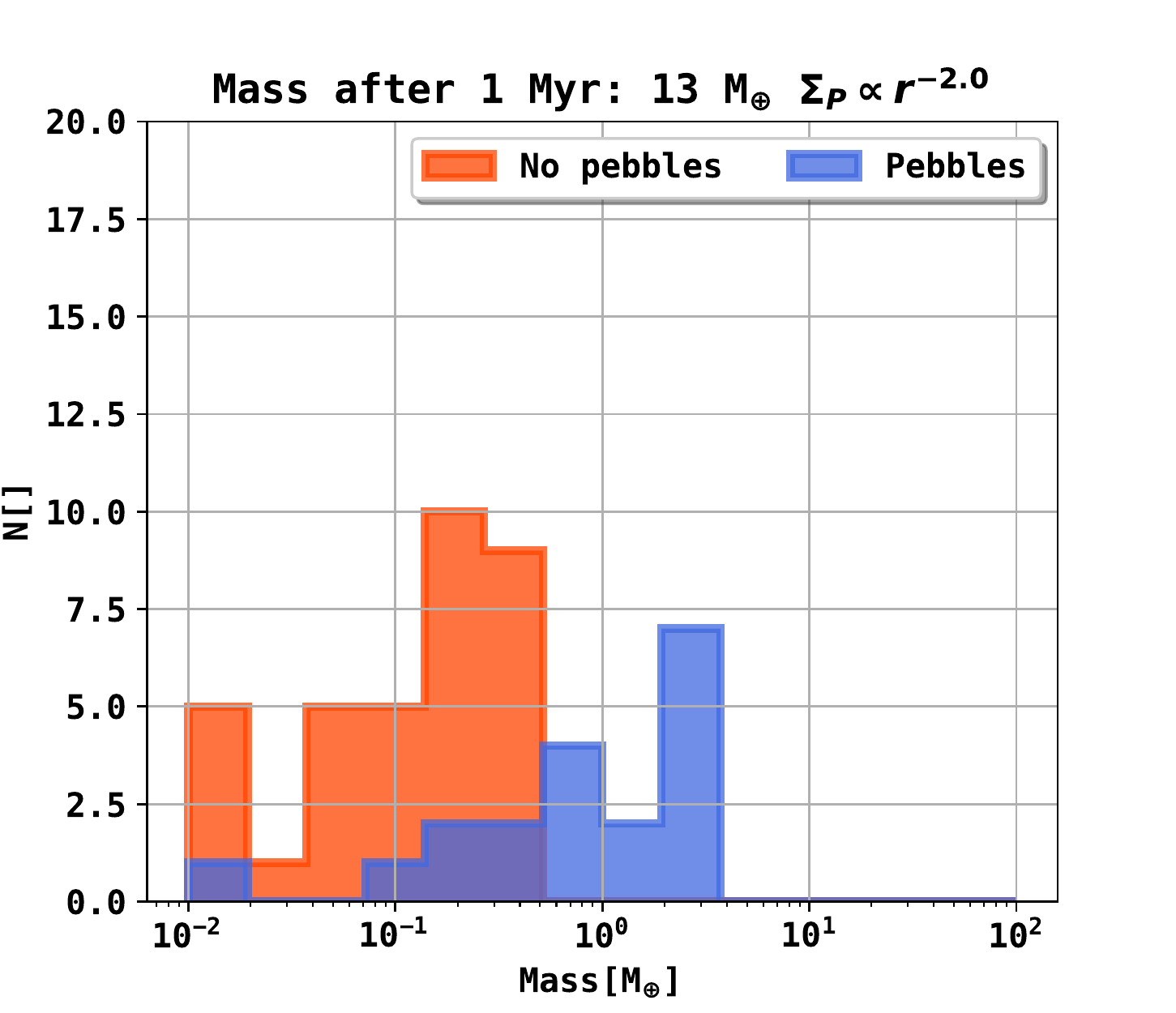}
\end{minipage}%
\\
\begin{minipage}{.33\textwidth}
  \centering
  \includegraphics[width=1.0\linewidth]{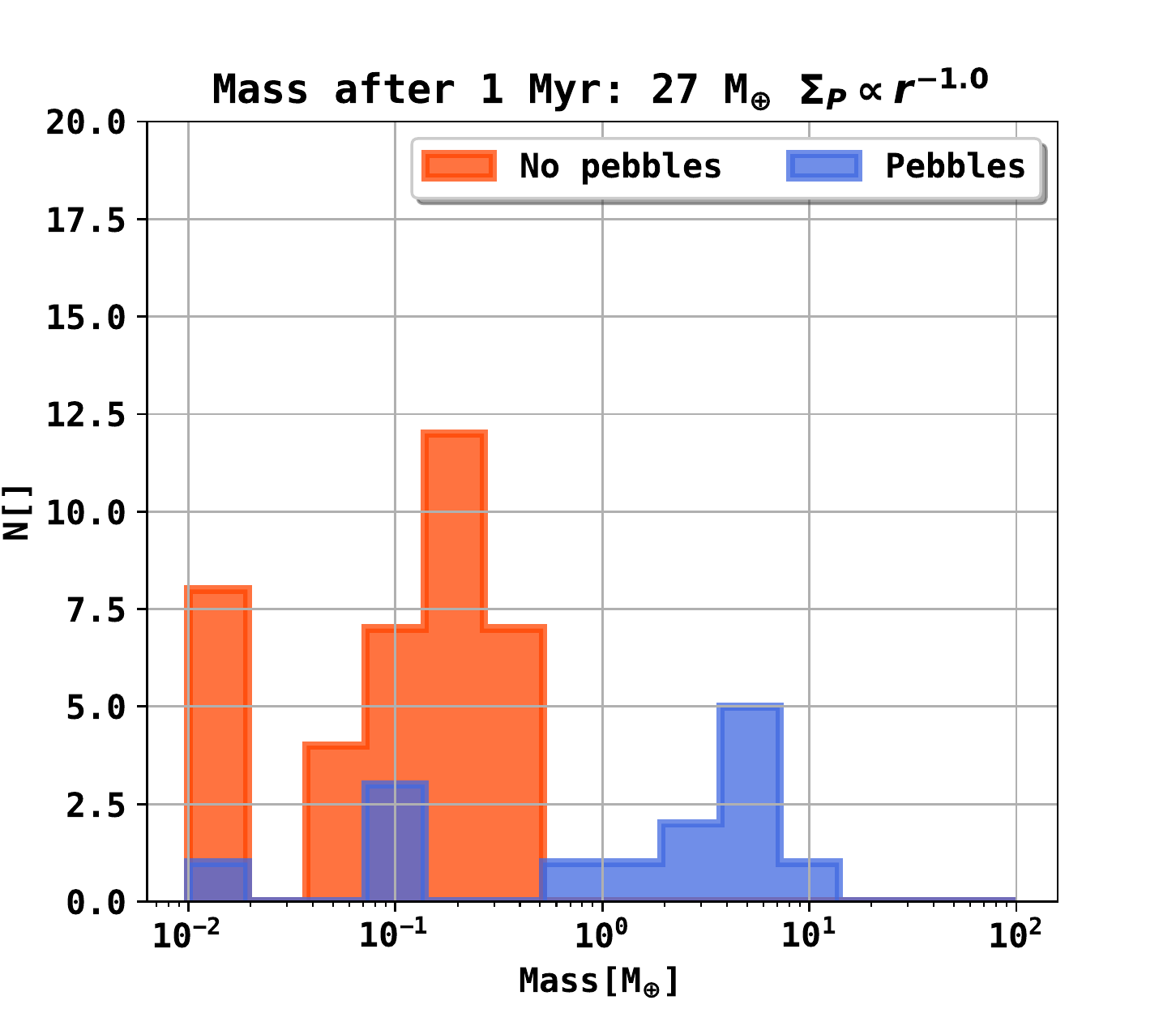}
\end{minipage}
\begin{minipage}{.33\textwidth}
  \includegraphics[width=1.0\linewidth]{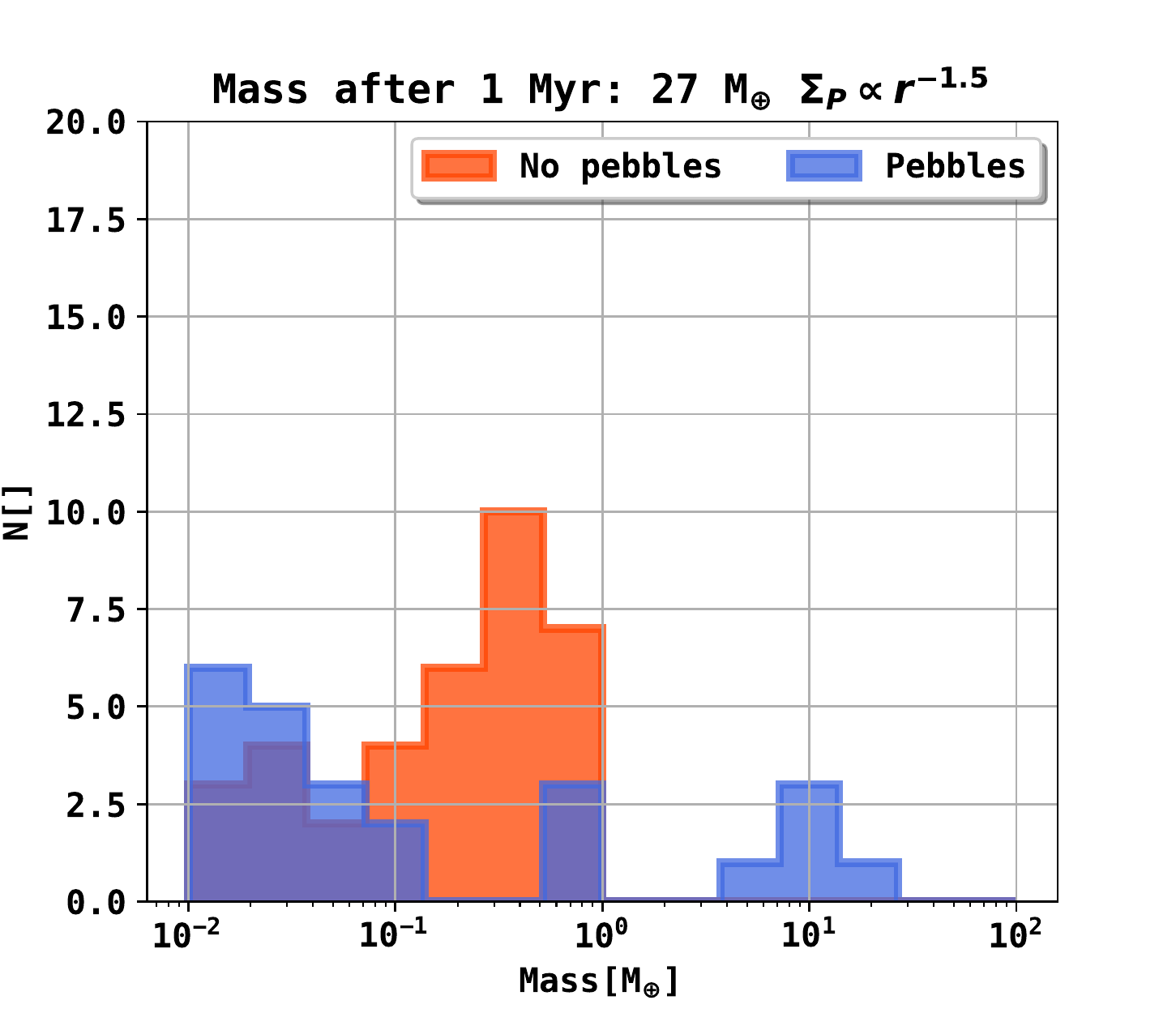}
\end{minipage}%
\begin{minipage}{.33\textwidth}
  \includegraphics[width=1.0\linewidth]{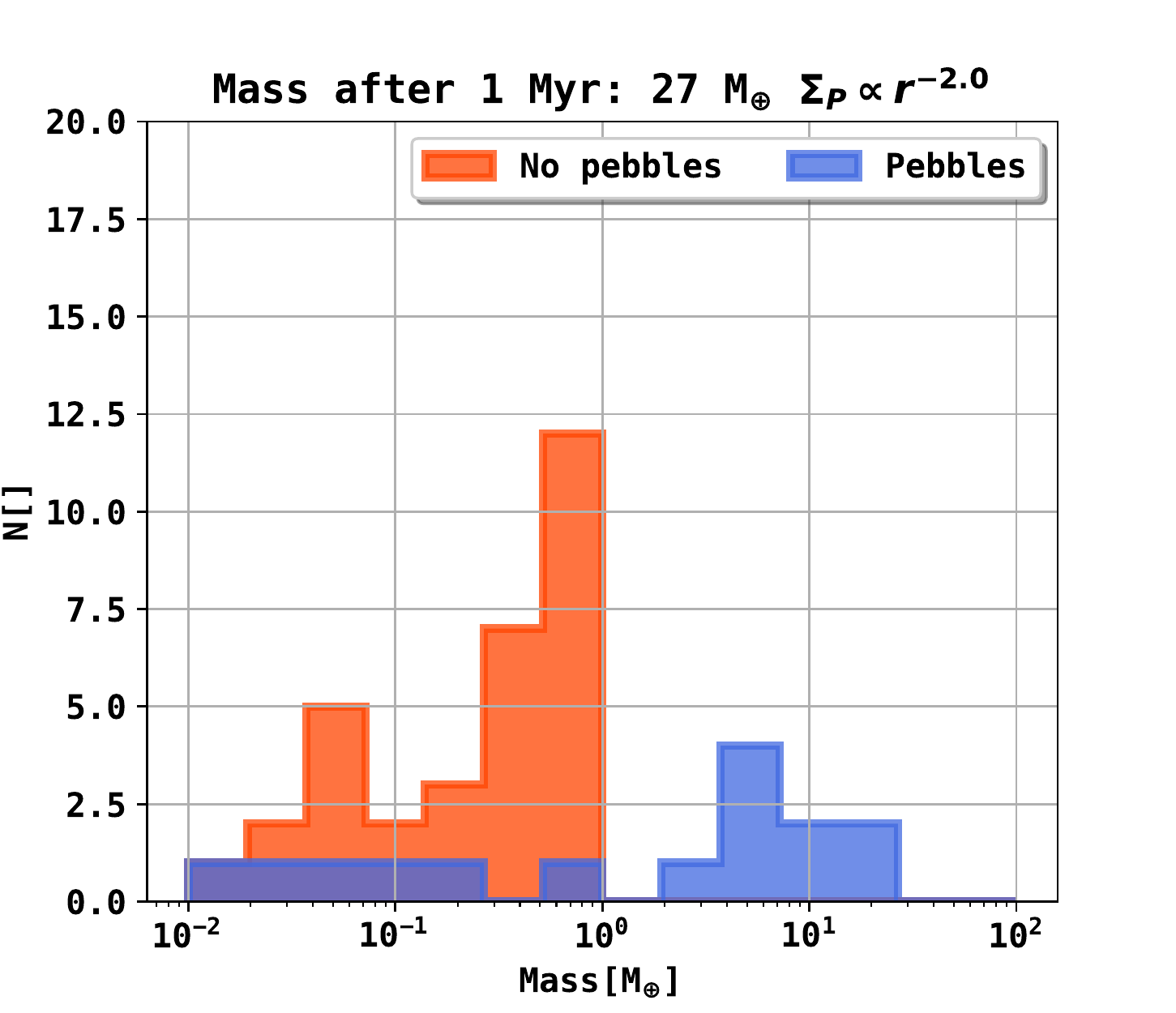}
\end{minipage}%
\caption{\small Embryo masses after 1 Myrs for the different parameters from Fig. \ref{Fig:Emb_form_LIPAD_6_ME} - Fig. \ref{Fig:Emb_form_LIPAD_27_ME}. The orange histograms show the systems in which pebble accretion is neglected, whereas the blue histograms show the systems in which pebble accretion is enabled.
}
\label{Fig:Embryo_masses}
\end{figure*}

Fig. \ref{Fig:Embryo_masses} shows the number of different embryo masses after 1 Myrs for the systems from Fig. \ref{Fig:Emb_form_LIPAD_6_ME} - Fig. \ref{Fig:Emb_form_LIPAD_27_ME}. The blue and orange histograms refer to simulations where we considered and not considered pebble accretion, respectively.  We see that without pebble accretion, there is no embryo with a mass higher than  1$\,$M$_{\oplus}$, whereas this is a very common outcome for the simulations in which pebble accretion is included. Generally in every system, the highest mass is achieved when pebble accretion is included.
\\
While the systems in which pebble accretion is neglected fail to build super earths with our input parameters, the formation of super earth planets becomes possible when including pebble accretion. While the number of active embryos decreases if pebble accretion is included ( see Fig. \ref{Fig:Embryo_number}), their masses increase drastically.
\subsection{Active number and total mass}
\label{Subsec:Active_number}
\begin{figure*}[]
\centering
\begin{minipage}{.33\textwidth}
  \centering
  \includegraphics[width=1.0\linewidth]{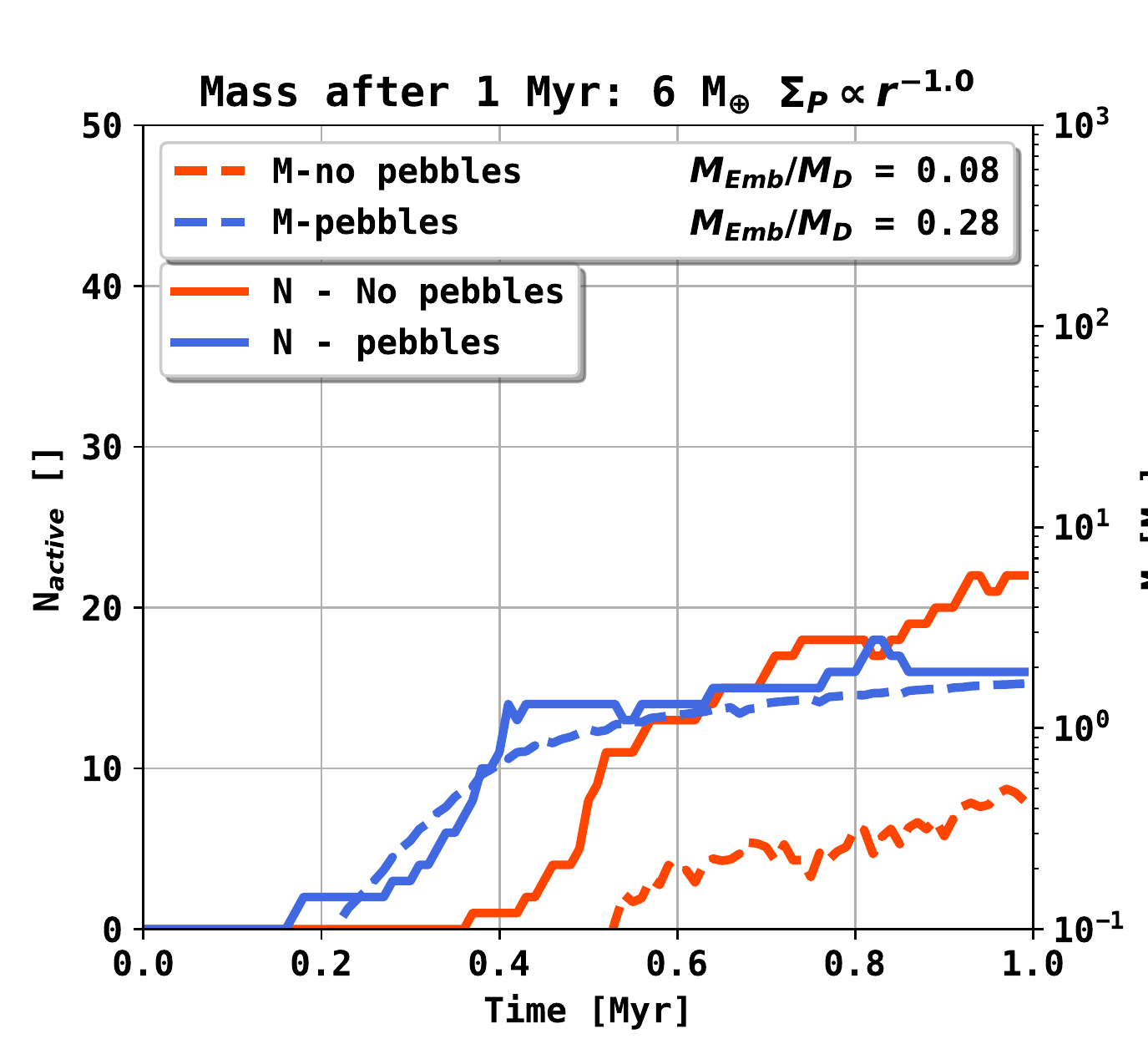}
\end{minipage}
\begin{minipage}{.33\textwidth}
  \includegraphics[width=1.0\linewidth]{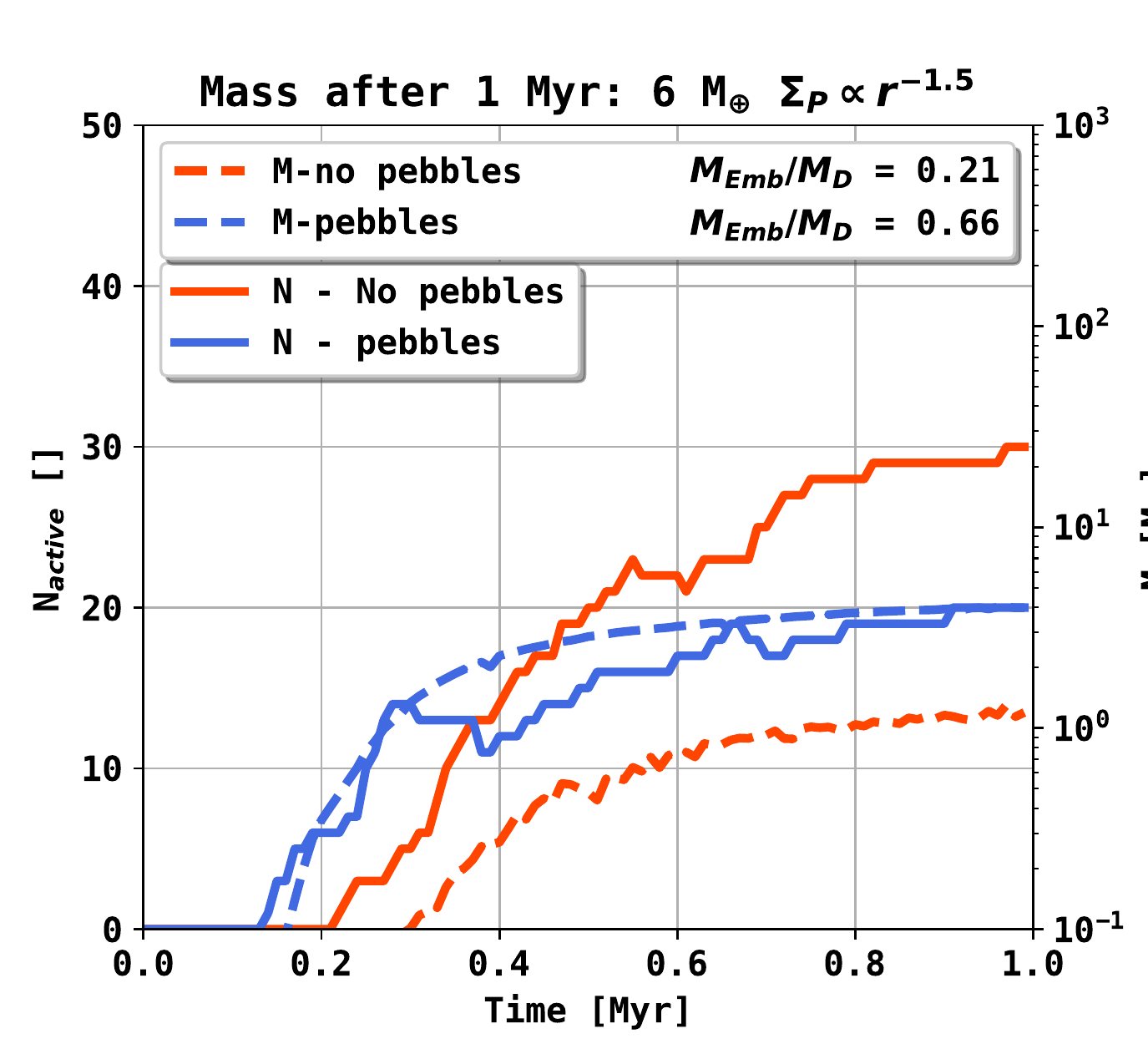}
\end{minipage}%
\begin{minipage}{.33\textwidth}
  \includegraphics[width=1.0\linewidth]{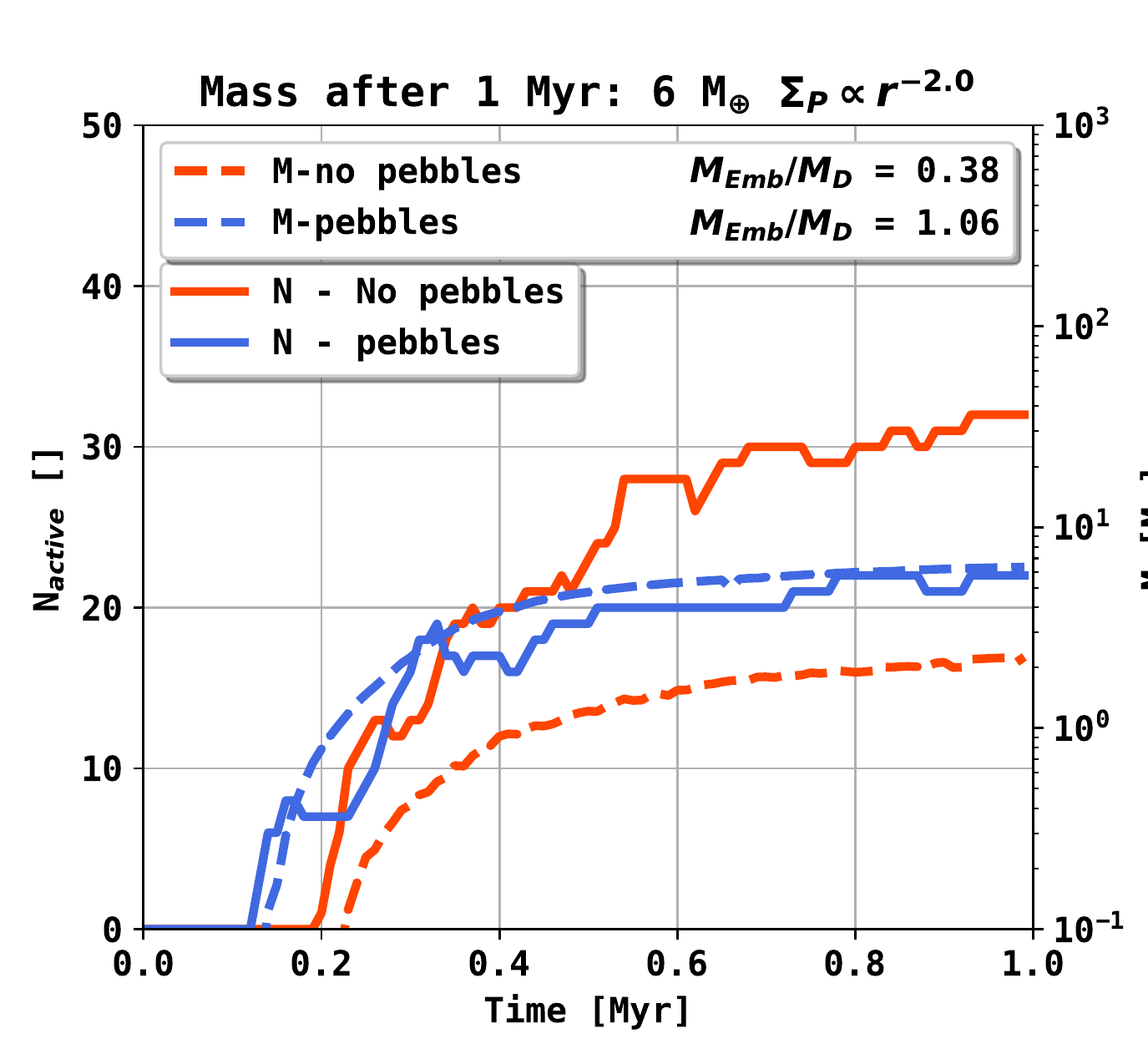}
\end{minipage}%
\\
\begin{minipage}{.33\textwidth}
  \centering
  \includegraphics[width=1.0\linewidth]{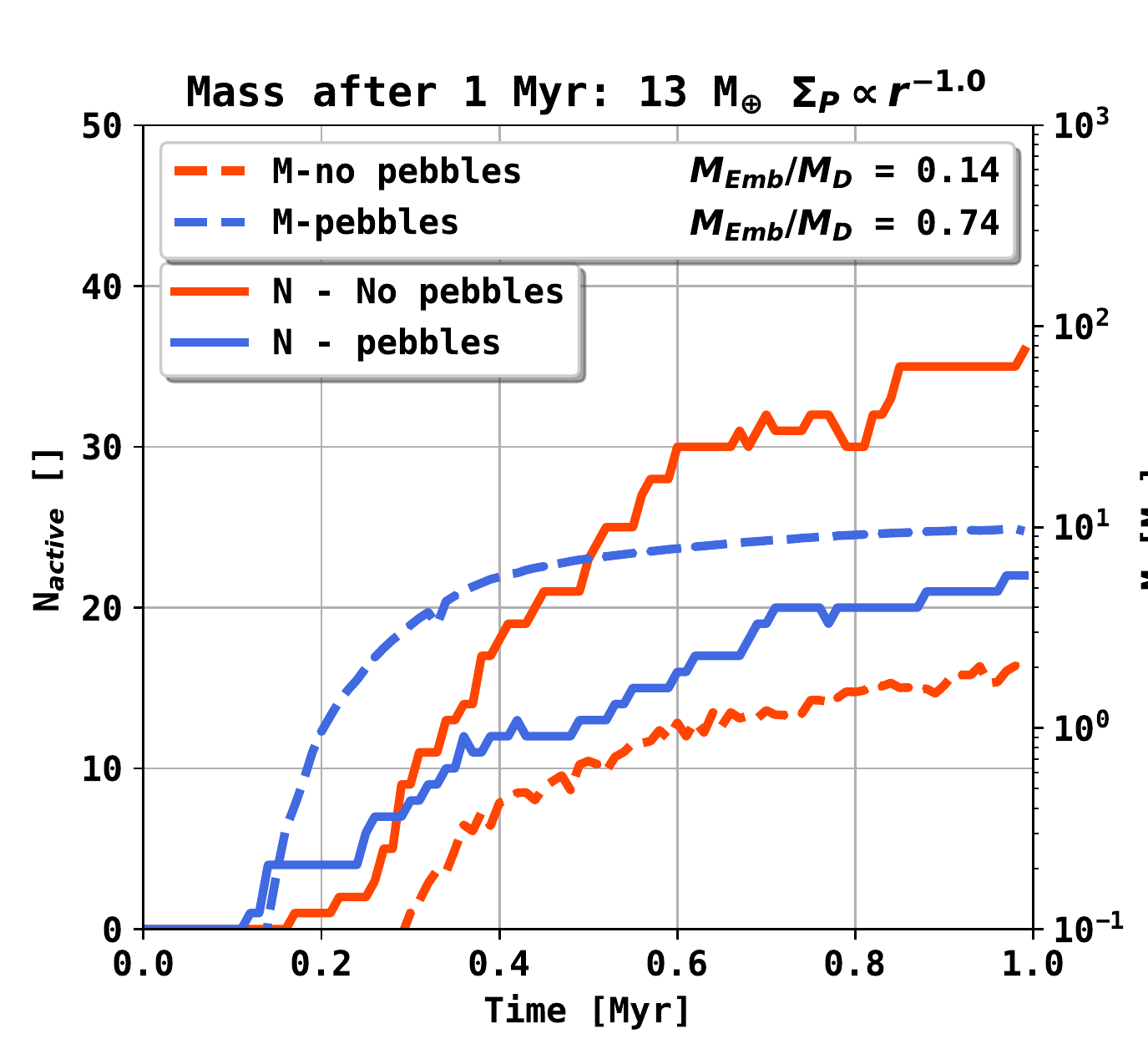}
\end{minipage}
\begin{minipage}{.33\textwidth}
  \includegraphics[width=1.0\linewidth]{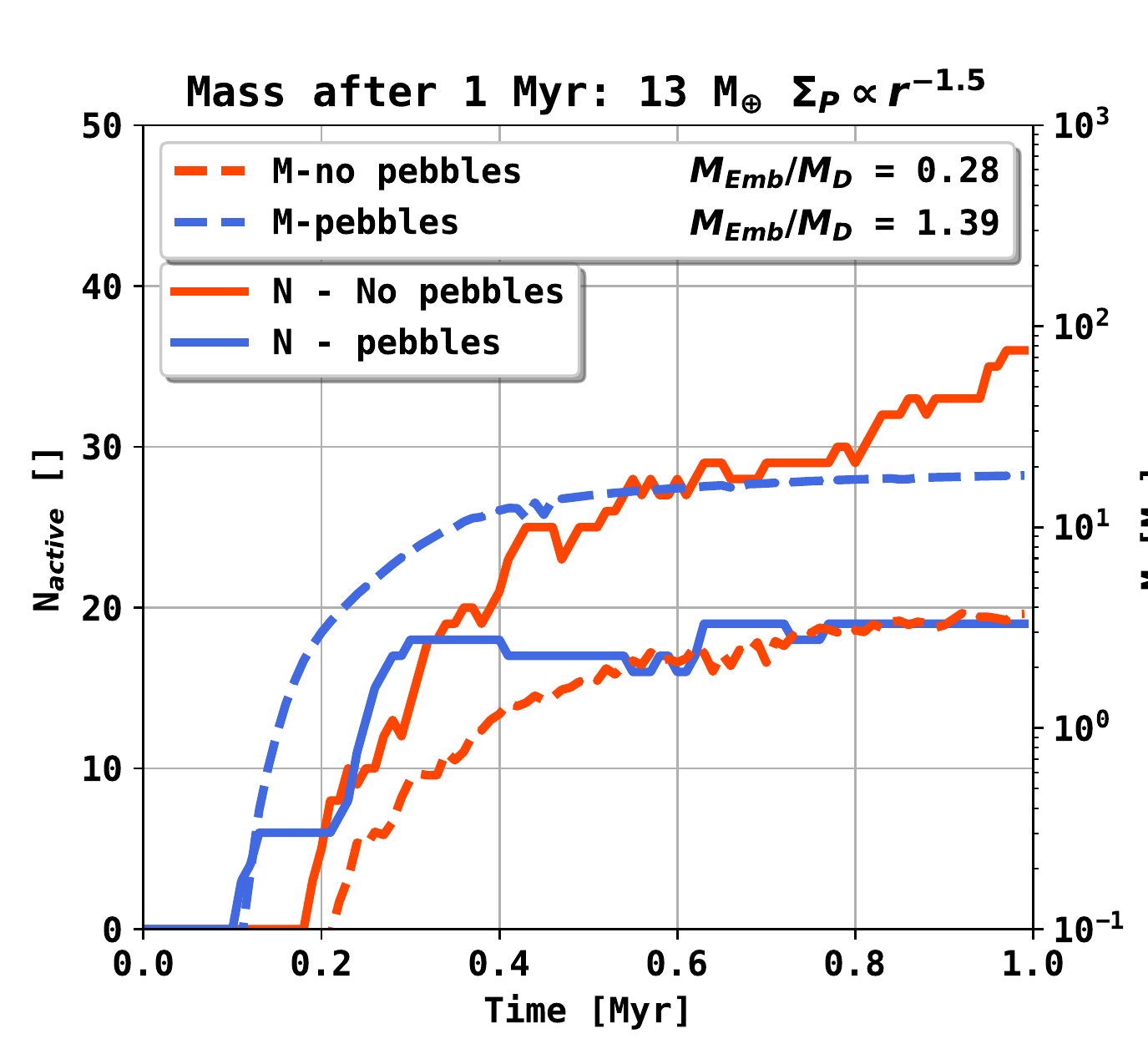}
\end{minipage}%
\begin{minipage}{.33\textwidth}
  \includegraphics[width=1.0\linewidth]{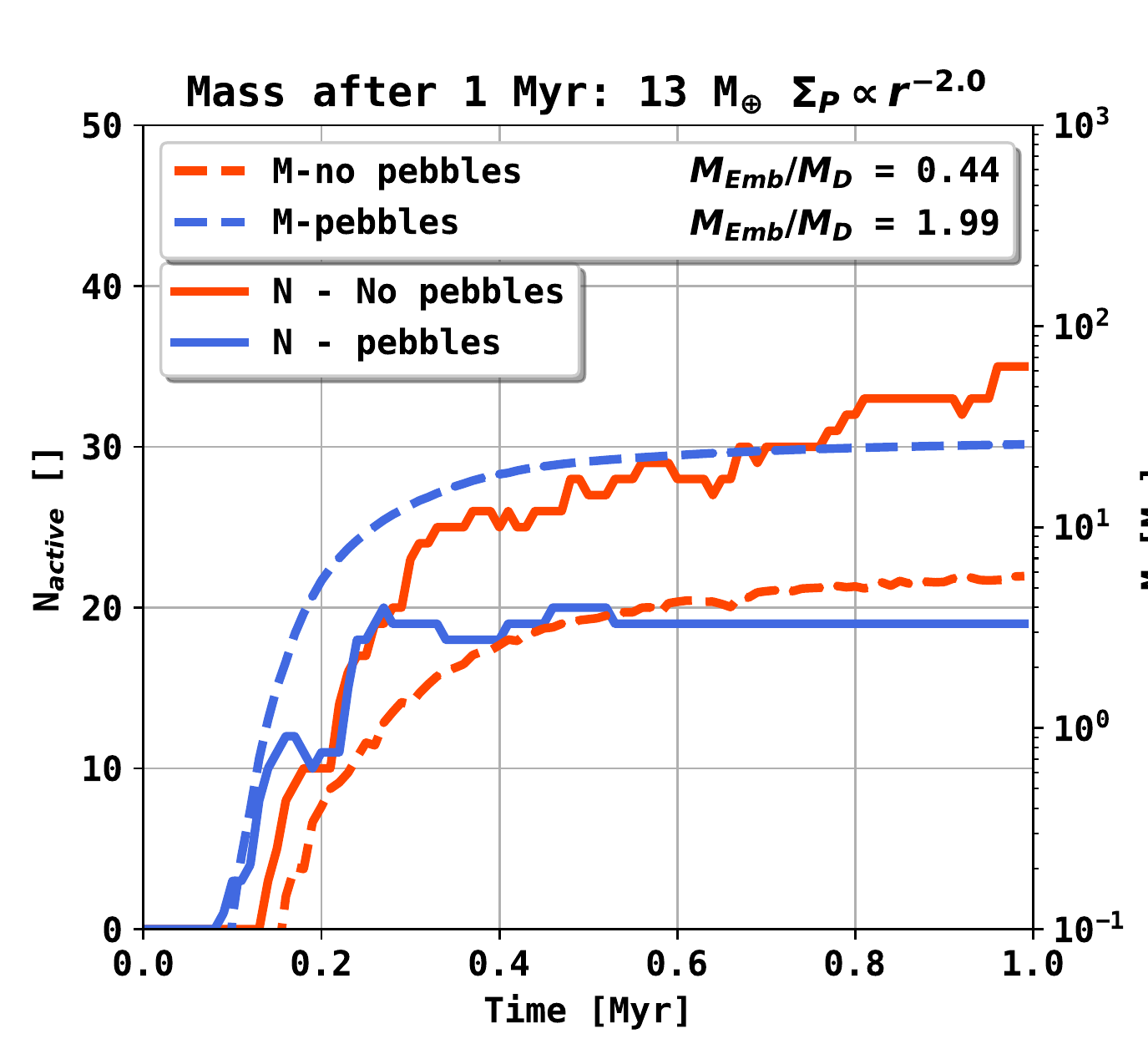}
\end{minipage}%
\\
\begin{minipage}{.33\textwidth}
  \centering
  \includegraphics[width=1.0\linewidth]{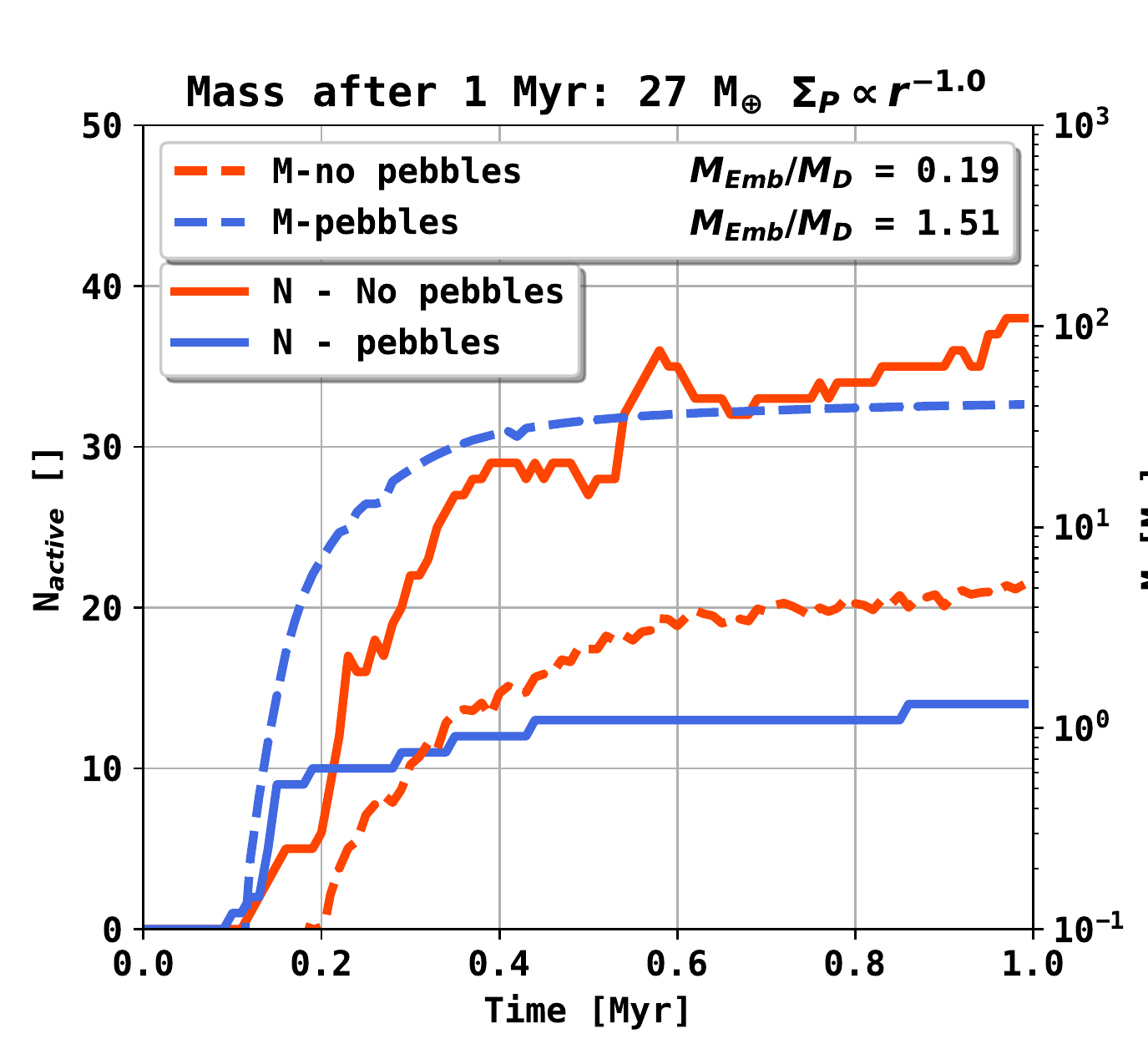}
\end{minipage}
\begin{minipage}{.33\textwidth}
  \includegraphics[width=1.0\linewidth]{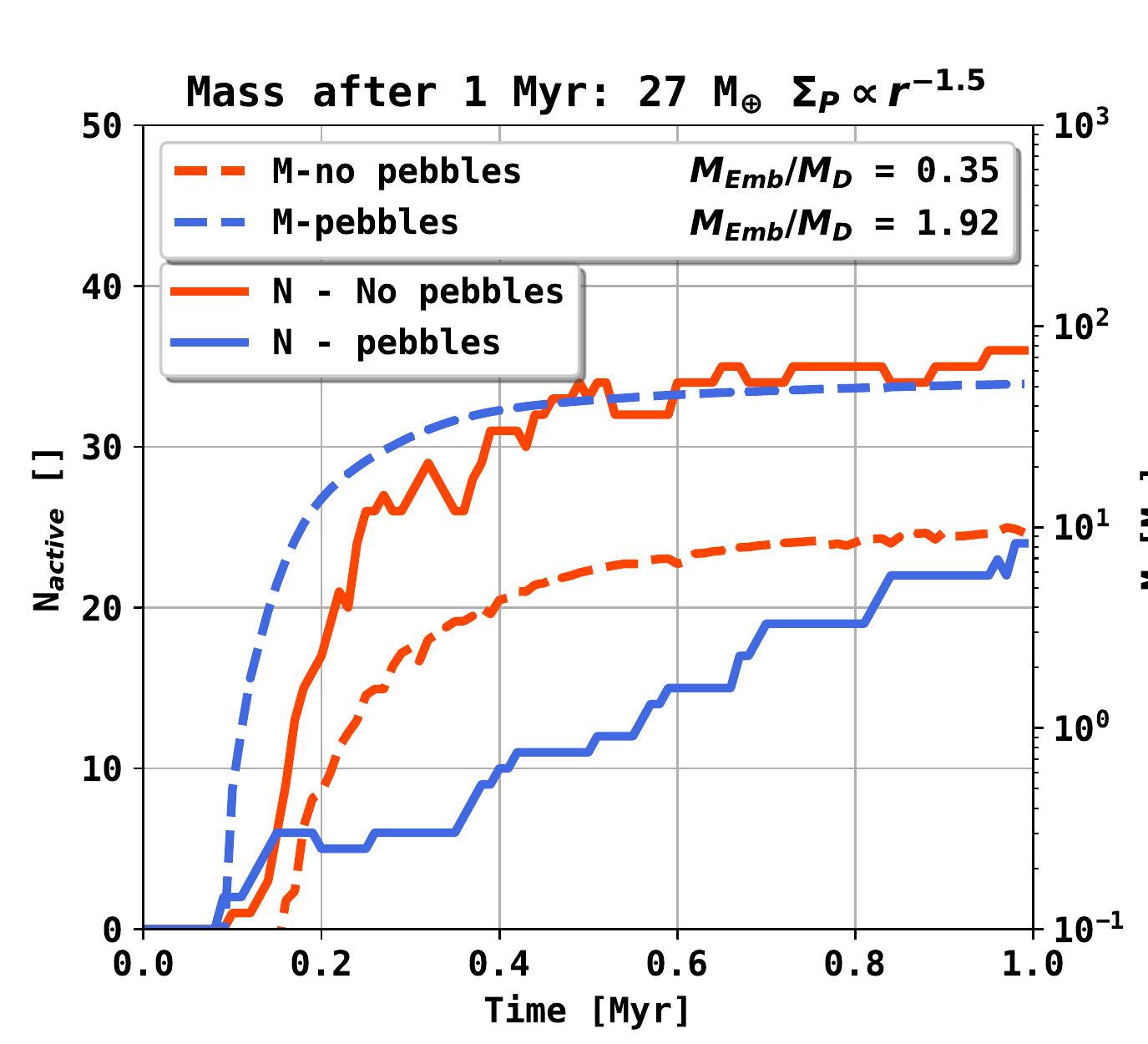}
\end{minipage}%
\begin{minipage}{.33\textwidth}
  \includegraphics[width=1.0\linewidth]{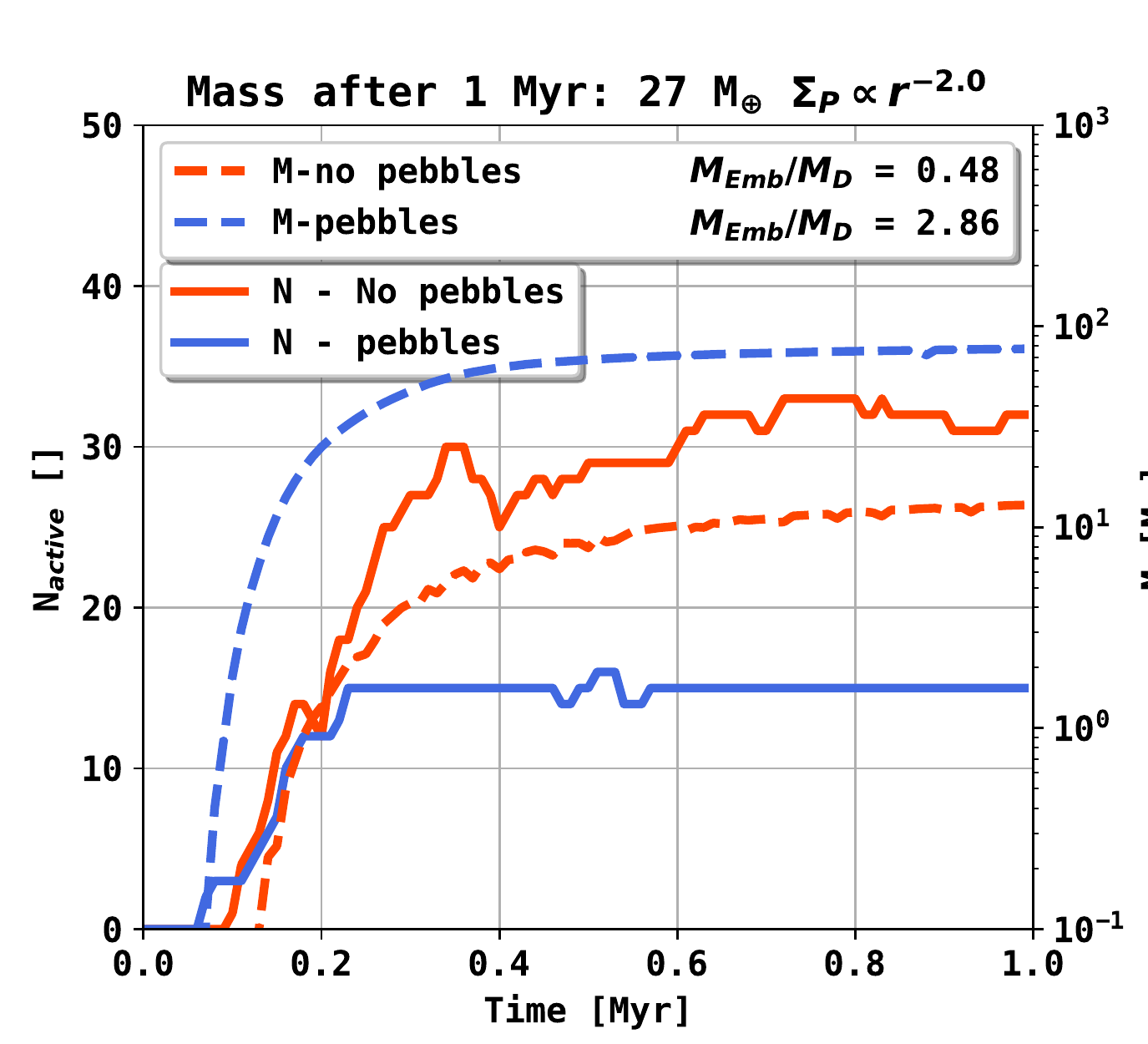}
\end{minipage}%
\caption{\small Number of active embryos (solid line) and total mass in embryos (dashed line) over time for the systems from Fig. \ref{Fig:Emb_form_LIPAD_6_ME} - Fig. \ref{Fig:Emb_form_LIPAD_27_ME}. The orange curves refer to the systems in which pebble accretion is disabled, whereas the blue lines refer to the systems in which pebble accretion is enabled. We also give the fraction of embryo mass over the total mass that entered the planetesimal disk after 1 Myrs ($M_{Emb}/M_{D}$). 
}
\label{Fig:Embryo_number}
\end{figure*}

Fig. \ref{Fig:Embryo_number} shows the total number of embryos and the total mass that is in embryos over time for the setups from Fig. \ref{Fig:Emb_form_LIPAD_6_ME} - Fig. \ref{Fig:Emb_form_LIPAD_27_ME}. We also give the fraction of total embryo mass M$_{Emb}$ over the mass that was given to the planetesimal disk after 1 Myr (M$_{D}$) for each setup. The first embryos always form in the systems in which pebble accretion is enabled. However, the number of active embryos during the simulation is almost a factor of 2 below the number of embryos in the systems without pebble accretion. The mass in embryos differs even more strongly than the active number of embryos for the corresponding systems. The fraction M$_{Emb}/$M$_{D}$ consistently increases for higher total masses and steeper $\Sigma_P$-profiles respectively. In the systems in which pebble accretion is included, it can exceed unity. This means that the mass in planetary embryos can be higher than the mass that is transformed into planetesimals, due to pebble accretion.
\subsection{Orbital Separation}
\label{Subsec:Orbital_separation}
\begin{figure*}[]
\centering
\begin{minipage}{.33\textwidth}
  \centering
  \includegraphics[width=1.0\linewidth]{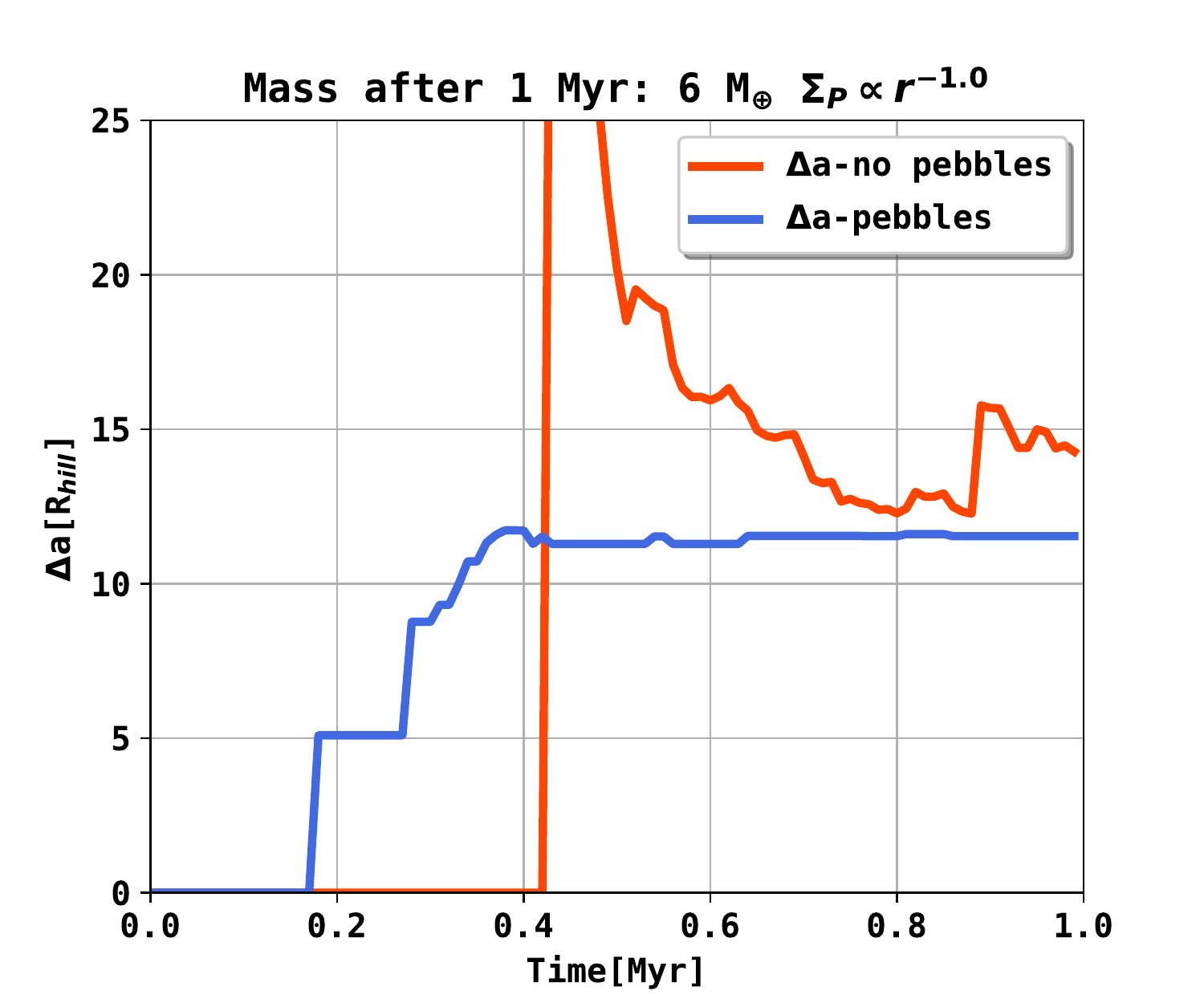}
\end{minipage}
\begin{minipage}{.33\textwidth}
  \includegraphics[width=1.0\linewidth]{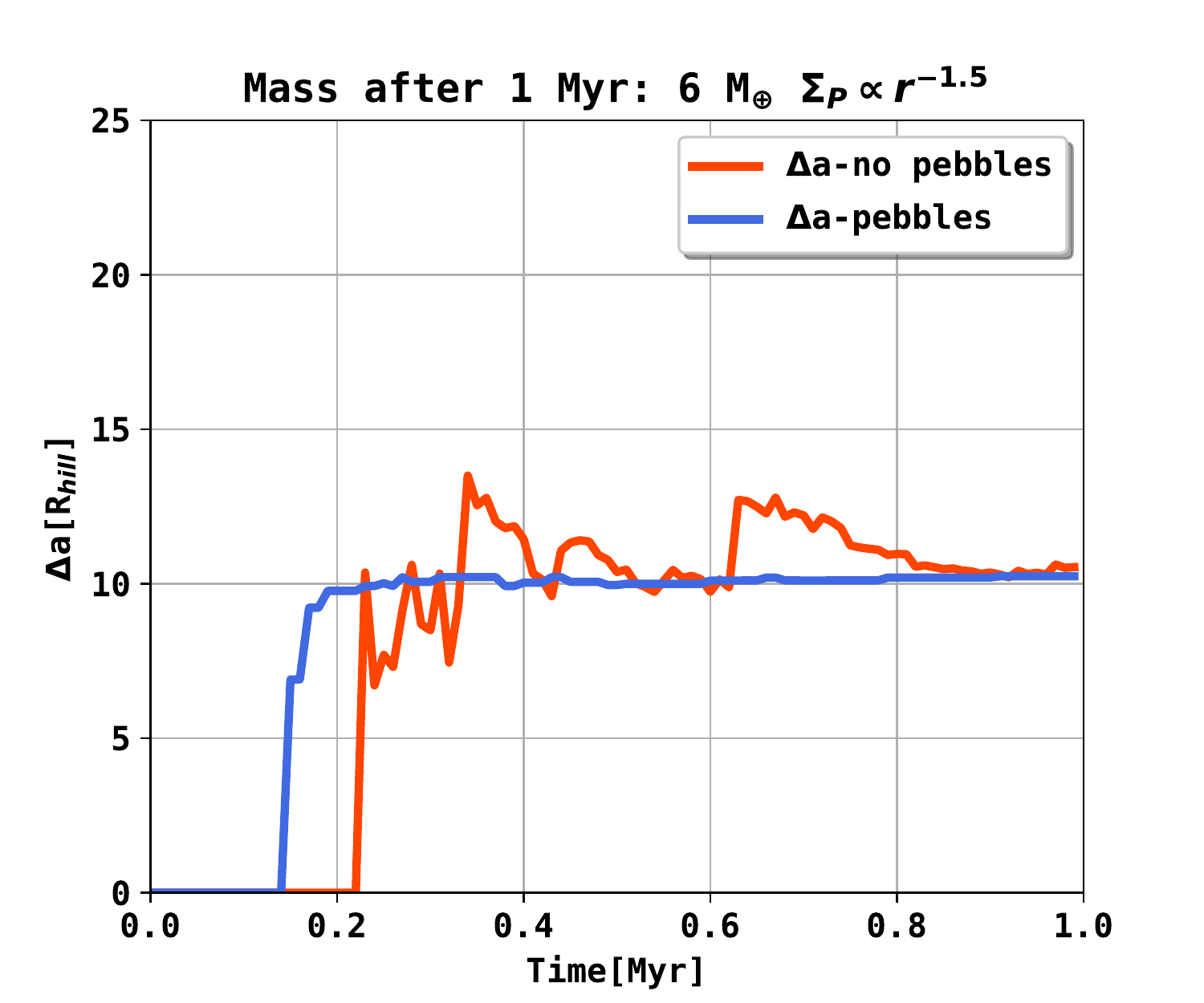}
\end{minipage}%
\begin{minipage}{.33\textwidth}
  \includegraphics[width=1.0\linewidth]{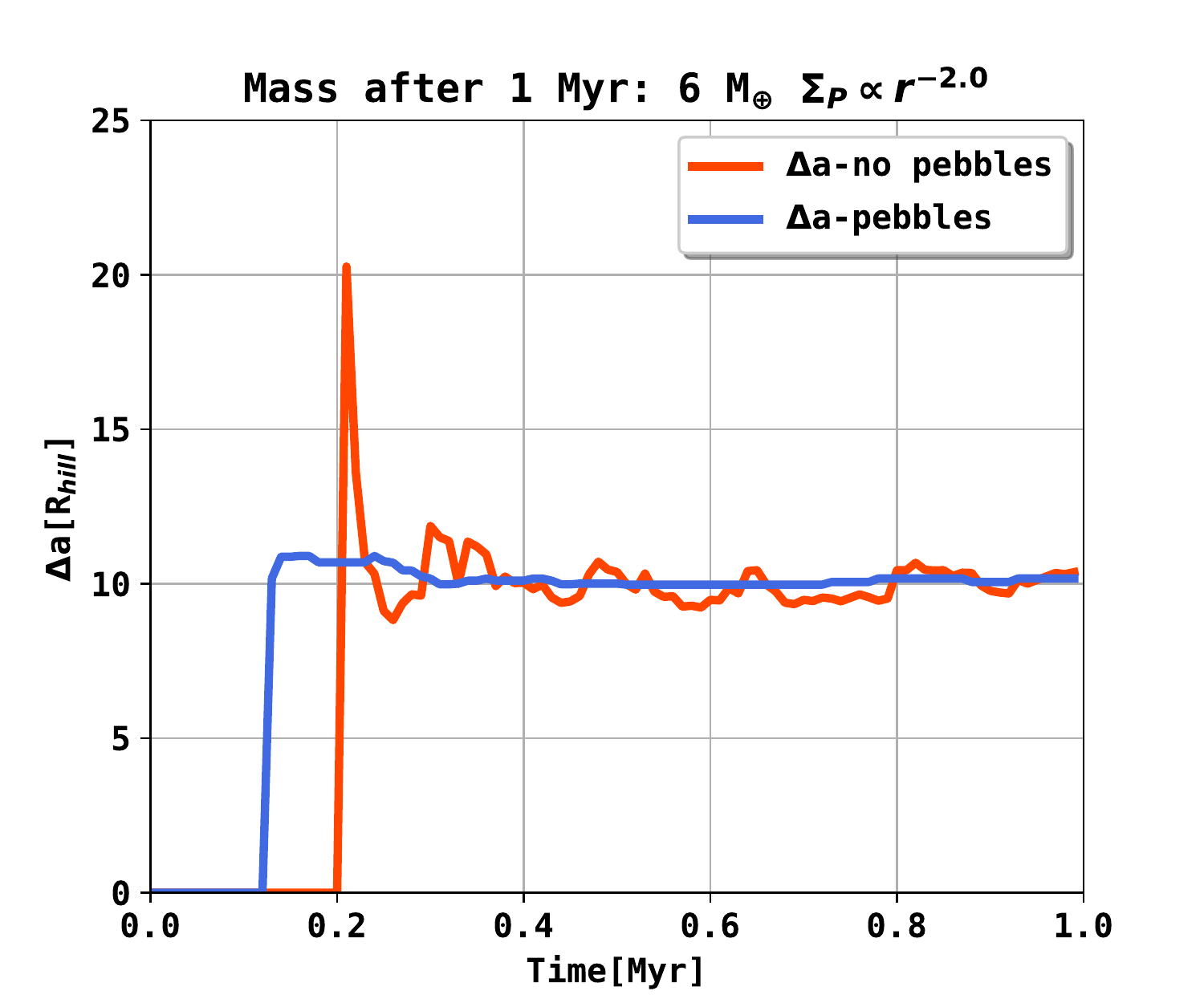}
\end{minipage}%
\\
\begin{minipage}{.33\textwidth}
  \centering
  \includegraphics[width=1.0\linewidth]{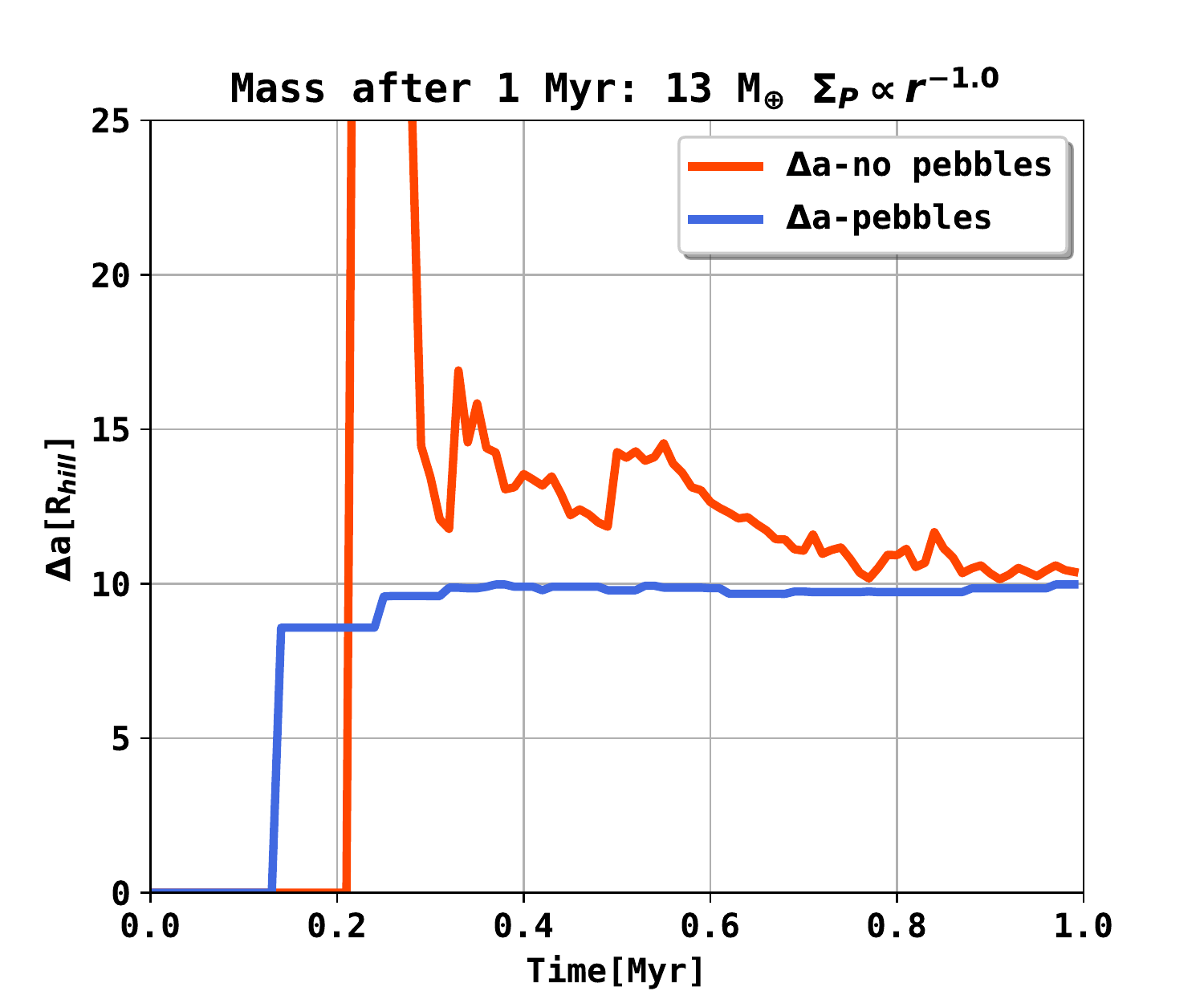}
\end{minipage}
\begin{minipage}{.33\textwidth}
  \includegraphics[width=1.0\linewidth]{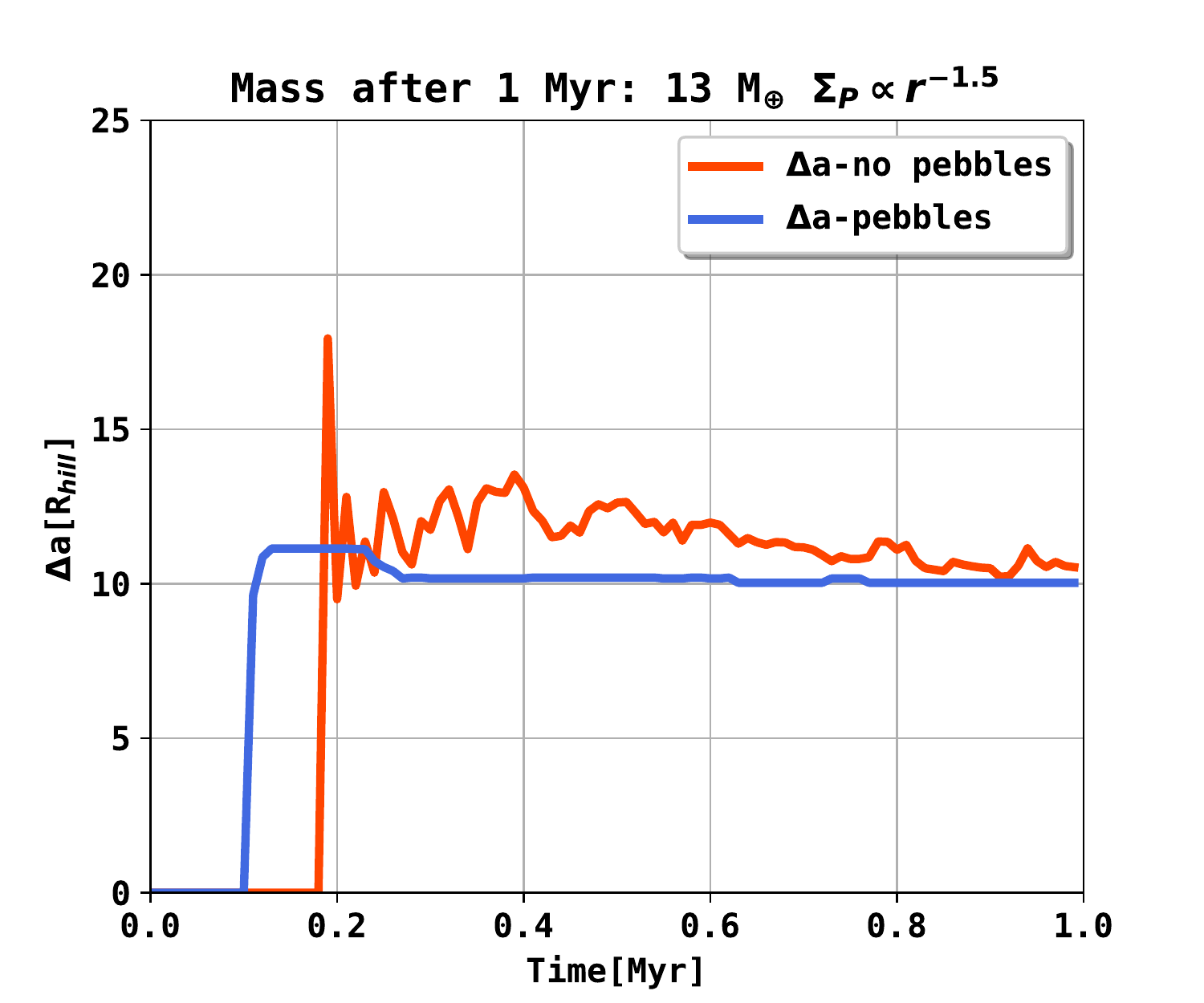}
\end{minipage}%
\begin{minipage}{.33\textwidth}
  \includegraphics[width=1.0\linewidth]{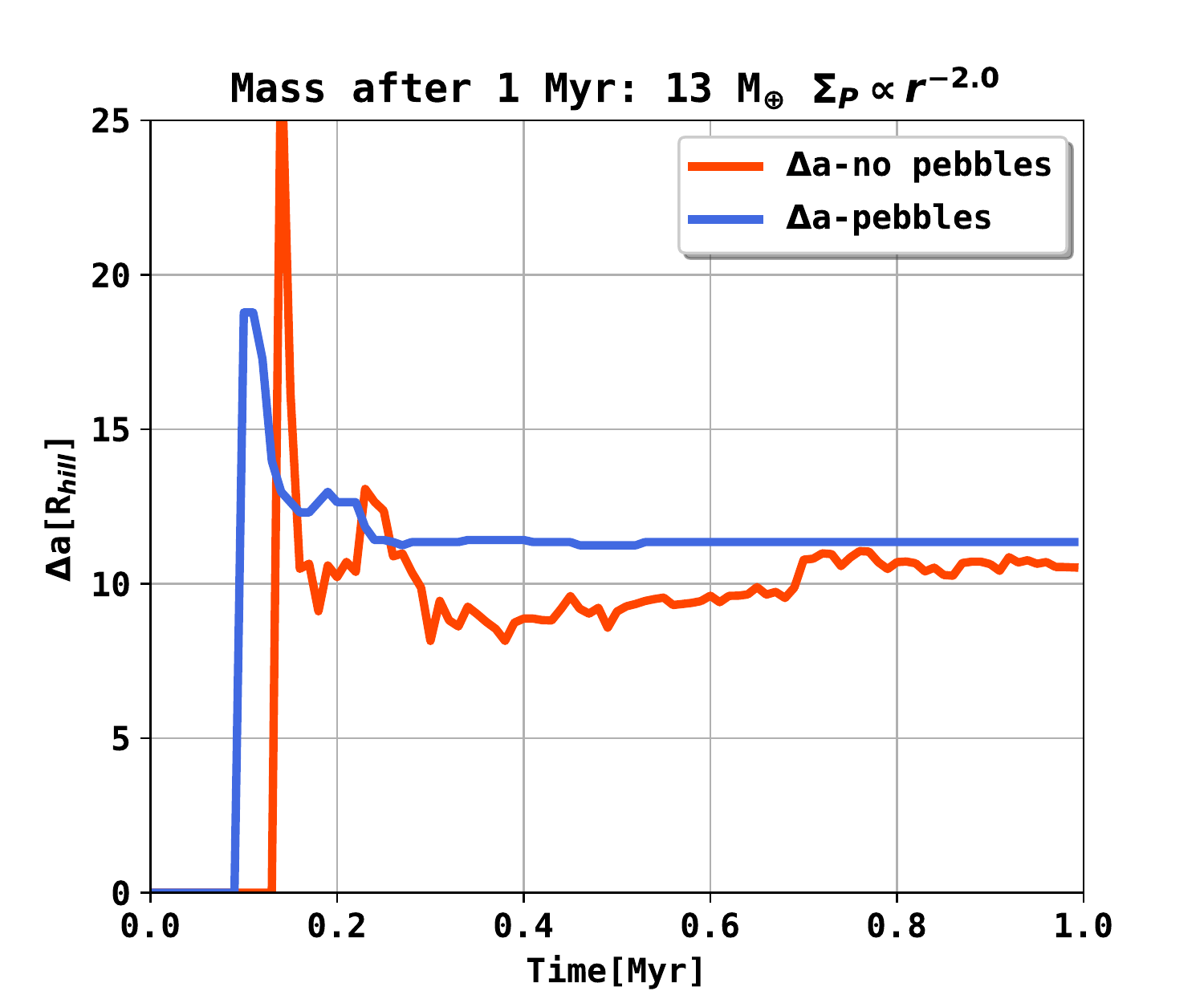}
\end{minipage}%
\\
\begin{minipage}{.33\textwidth}
  \centering
  \includegraphics[width=1.0\linewidth]{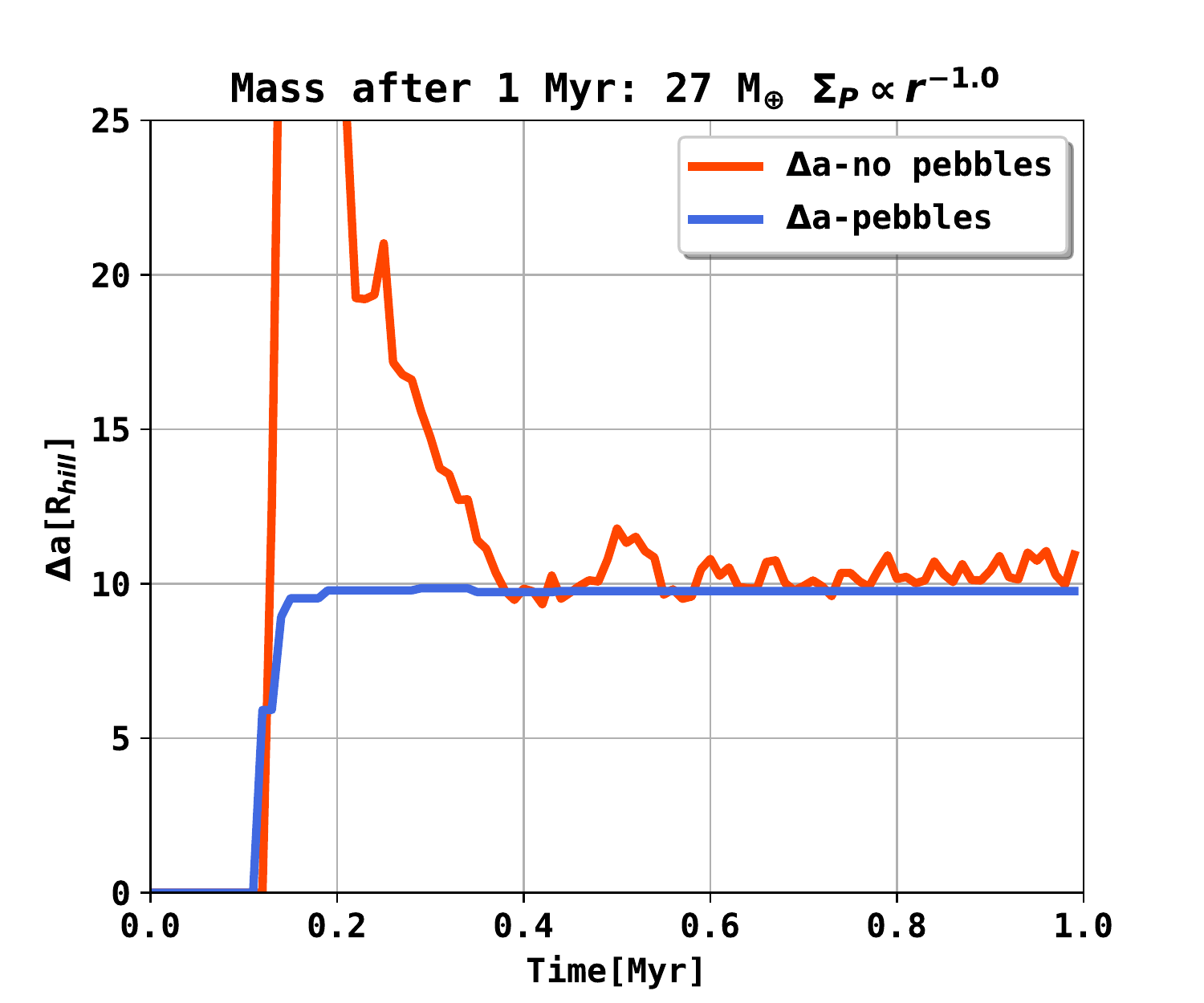}
\end{minipage}
\begin{minipage}{.33\textwidth}
  \includegraphics[width=1.0\linewidth]{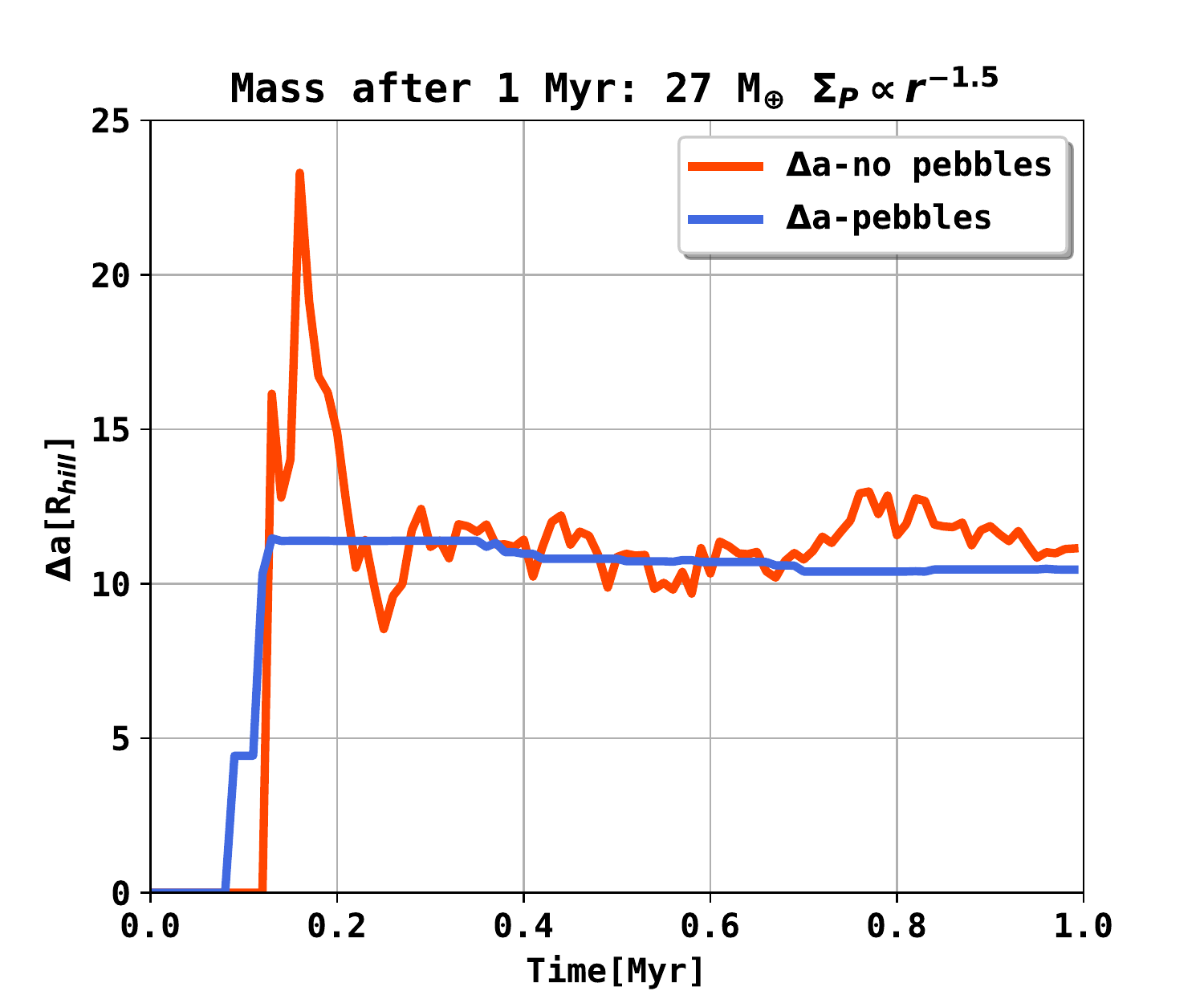}
\end{minipage}%
\begin{minipage}{.33\textwidth}
  \includegraphics[width=1.0\linewidth]{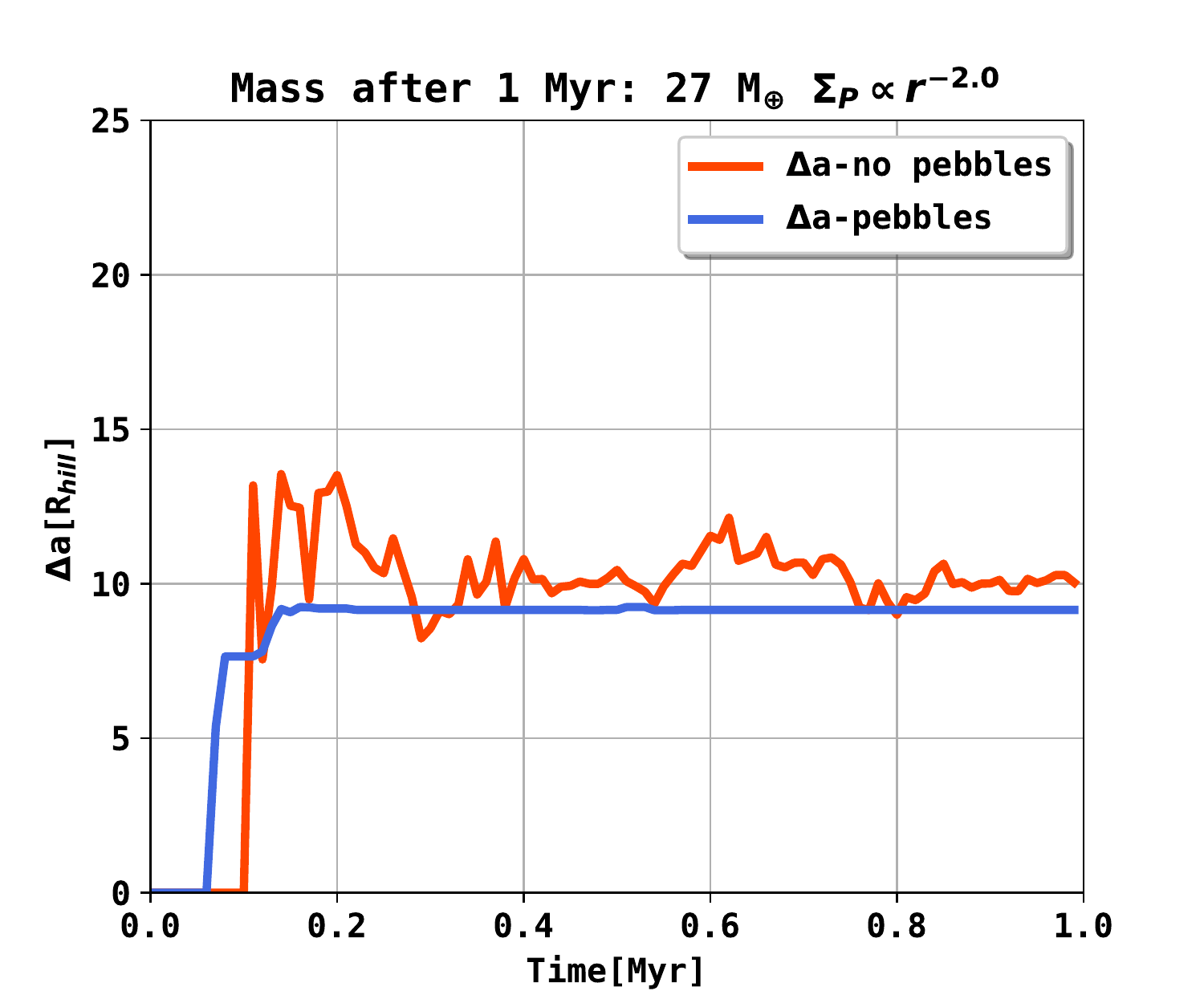}
\end{minipage}%
\caption{\small Orbital separation of active embryos over time from the systems from Fig. \ref{Fig:Emb_form_LIPAD_6_ME} - Fig. \ref{Fig:Emb_form_LIPAD_27_ME}. The orange curves refer to the systems in which pebble accretion is disabled, whereas the blue curves refer to the systems in which pebble accretion is enabled. The distance is expressed in units of the embryos Hill radii.
}
\label{Fig:Orbital_separation}
\end{figure*}

In Fig. \ref{Fig:Orbital_separation} we compare the mean orbital separation of embryos over time for the systems from Fig. \ref{Fig:Emb_form_LIPAD_6_ME} - Fig. \ref{Fig:Emb_form_LIPAD_27_ME}. The orbital separation is expressed in units of the embryos Hill radii. We can see that the mean orbital separation after 1 Myr converges to $\approx 10$R$_{Hill}$ for each setup. The simulations in which pebble accretion is included show a smoother and more stable behavior over time than the systems in which pebble accretion is neglected. The explanation for these differences lies in the fact that the first embryos can start growing further apart from each other in the runs that only consider planetesimal accretion.  Therefore, numerous embryos are needed in order to converge for a characteristic orbital Hill spacing.
\\
When considering pebble accretion, embryos tend to initially grow closer to each other. Connecting the orbital separation from Fig. \ref{Fig:Orbital_separation} with the embryo masses from Fig. \ref{Fig:Embryo_masses} and the time semimajor axis evolution from Fig. \ref{Fig:Emb_form_LIPAD_6_ME} - Fig. \ref{Fig:Emb_form_LIPAD_27_ME}, we can see that the absolute physical distance between embryos increases largely due to their mass increase and therefore their increasing Hill radius. 
\\
The dynamical separation of embryos when expressed in Hill radii does not change, their physcial separation as a consequence does. The possible area of embryo formation on the other hand does not enlarge if pebble accretion is included (see Fig. \ref{Fig:Emb_form_LIPAD_6_ME} - Fig. \ref{Fig:Emb_form_LIPAD_27_ME}). Since the orbital separation increases, the number of active embryos within the possible area of embryo formation decreases, as a consequence of their rapid growth by pebble accretion. 
\subsection{Cummulative distribution}
\label{Subsec:Cummulative_distribution}
\begin{figure*}[]
\centering
\begin{minipage}{.33\textwidth}
  \centering
  \includegraphics[width=1.0\linewidth]{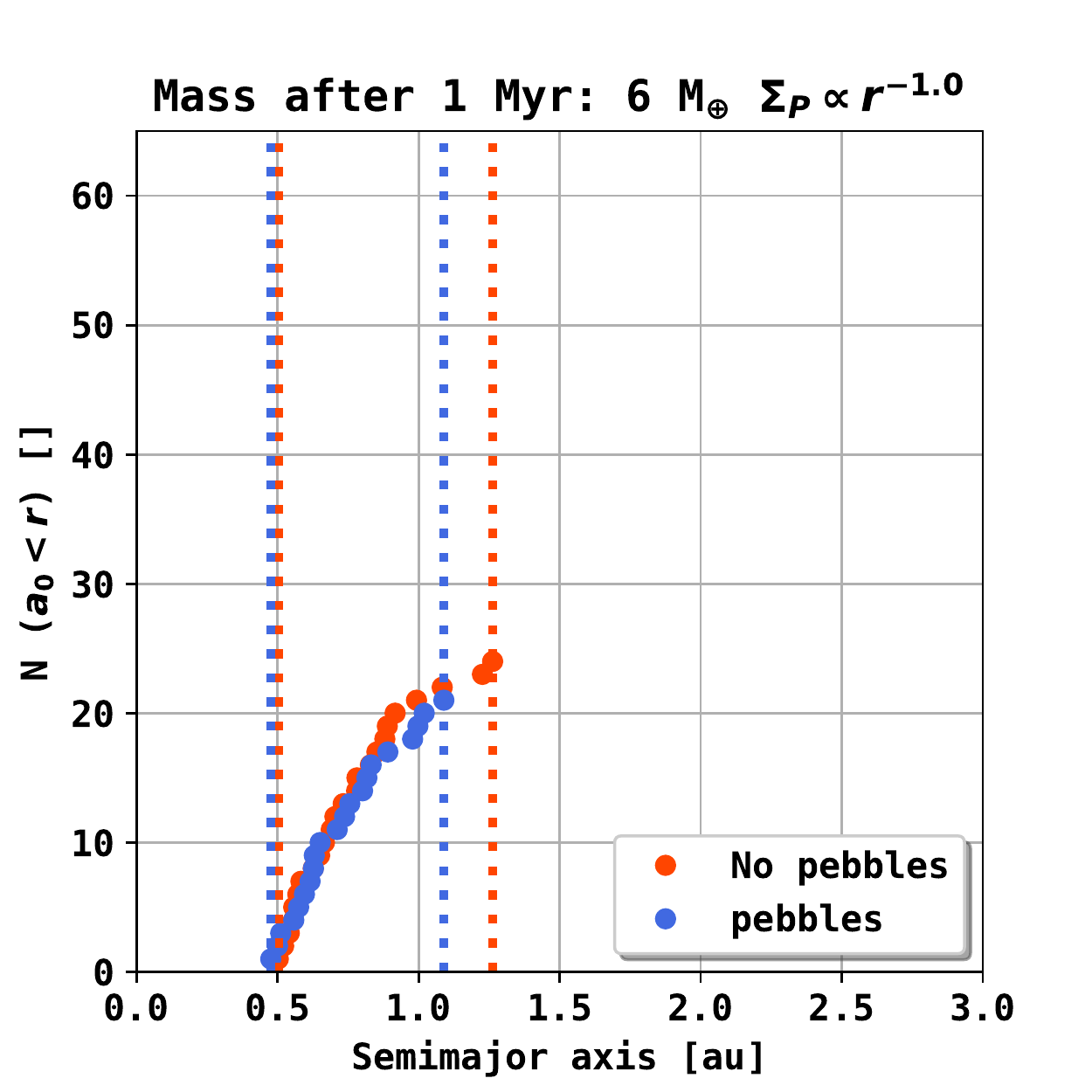}
\end{minipage}
\begin{minipage}{.33\textwidth}
  \includegraphics[width=1.0\linewidth]{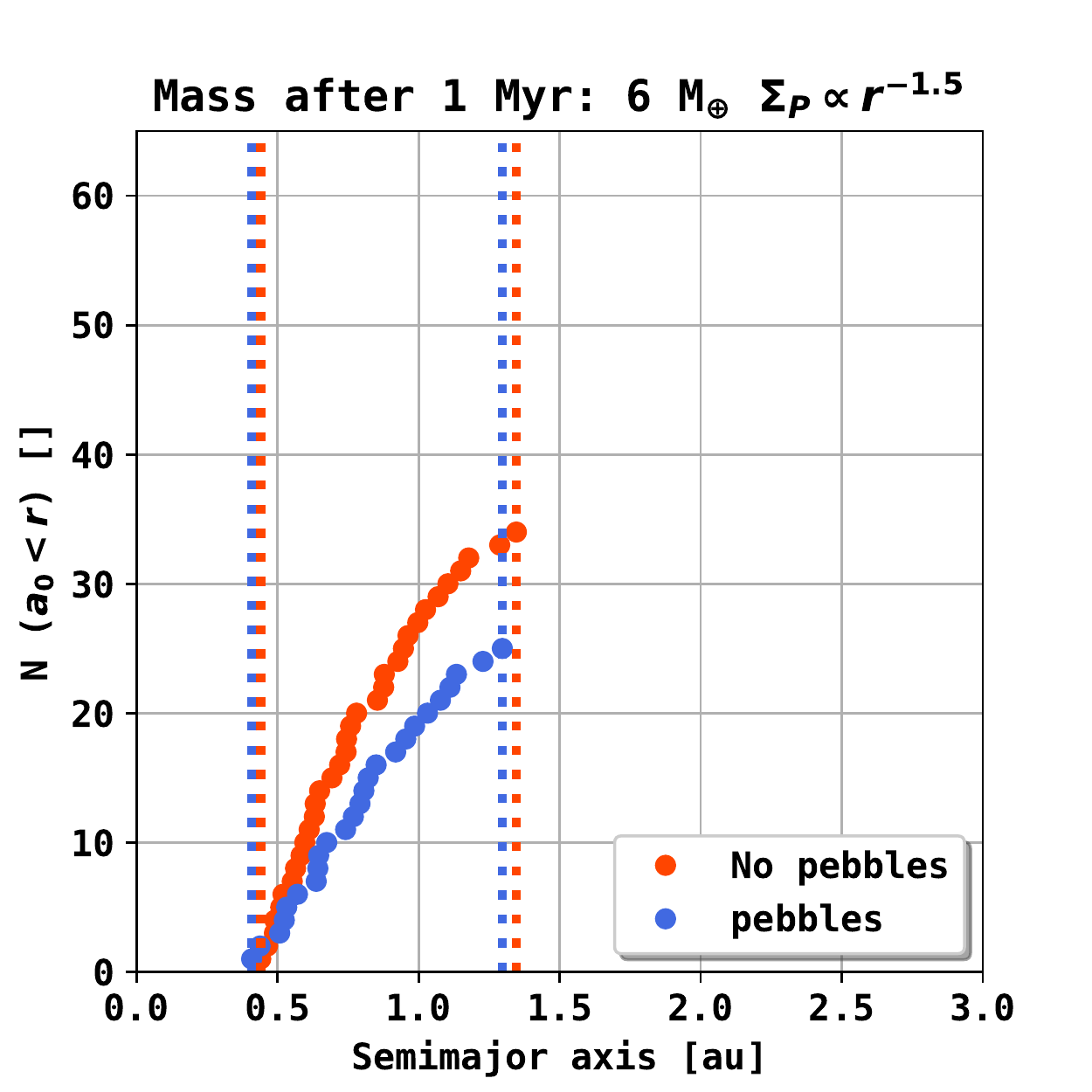}
\end{minipage}%
\begin{minipage}{.33\textwidth}
  \includegraphics[width=1.0\linewidth]{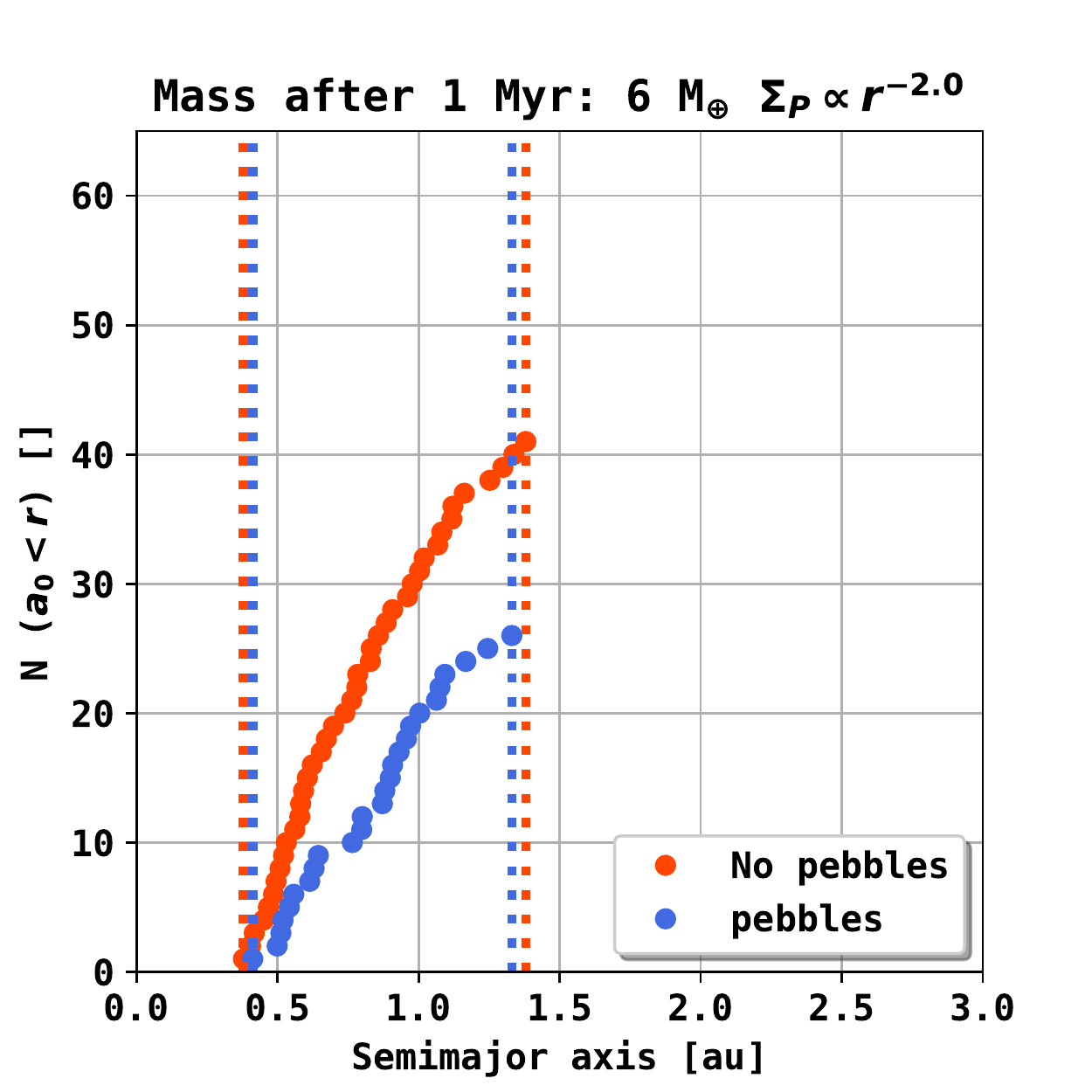}
\end{minipage}%
\\
\begin{minipage}{.33\textwidth}
  \centering
  \includegraphics[width=1.0\linewidth]{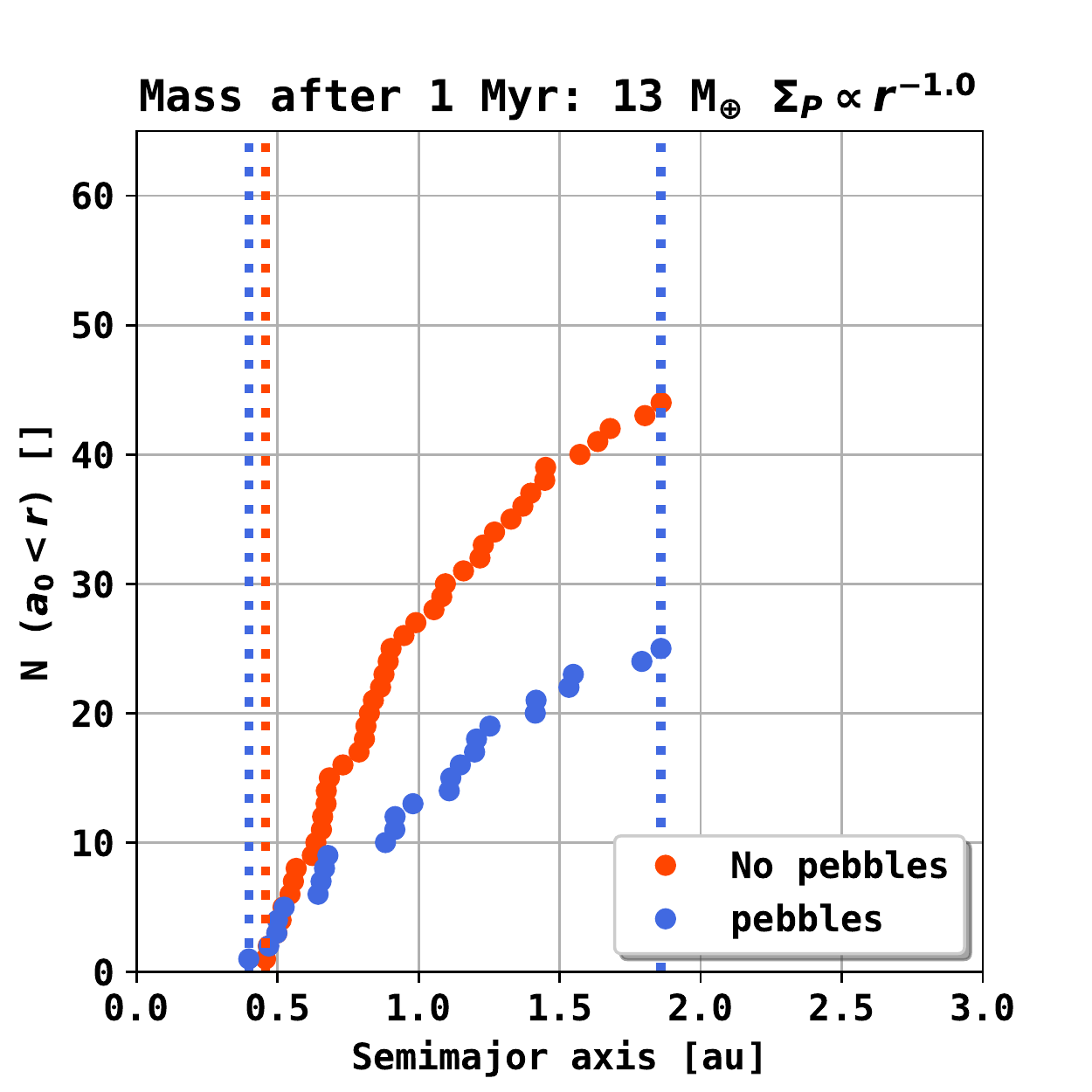}
\end{minipage}
\begin{minipage}{.33\textwidth}
  \includegraphics[width=1.0\linewidth]{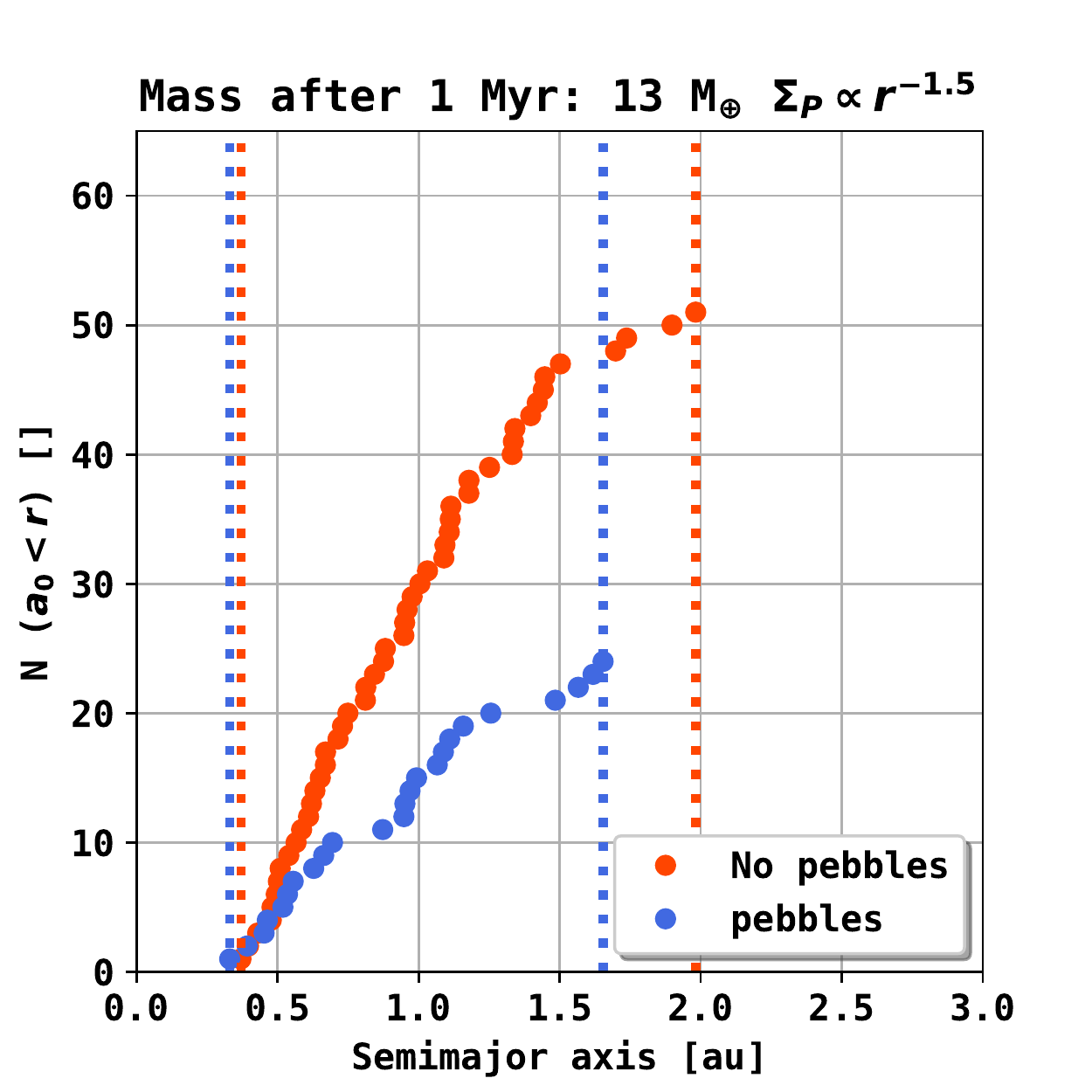}
\end{minipage}%
\begin{minipage}{.33\textwidth}
  \includegraphics[width=1.0\linewidth]{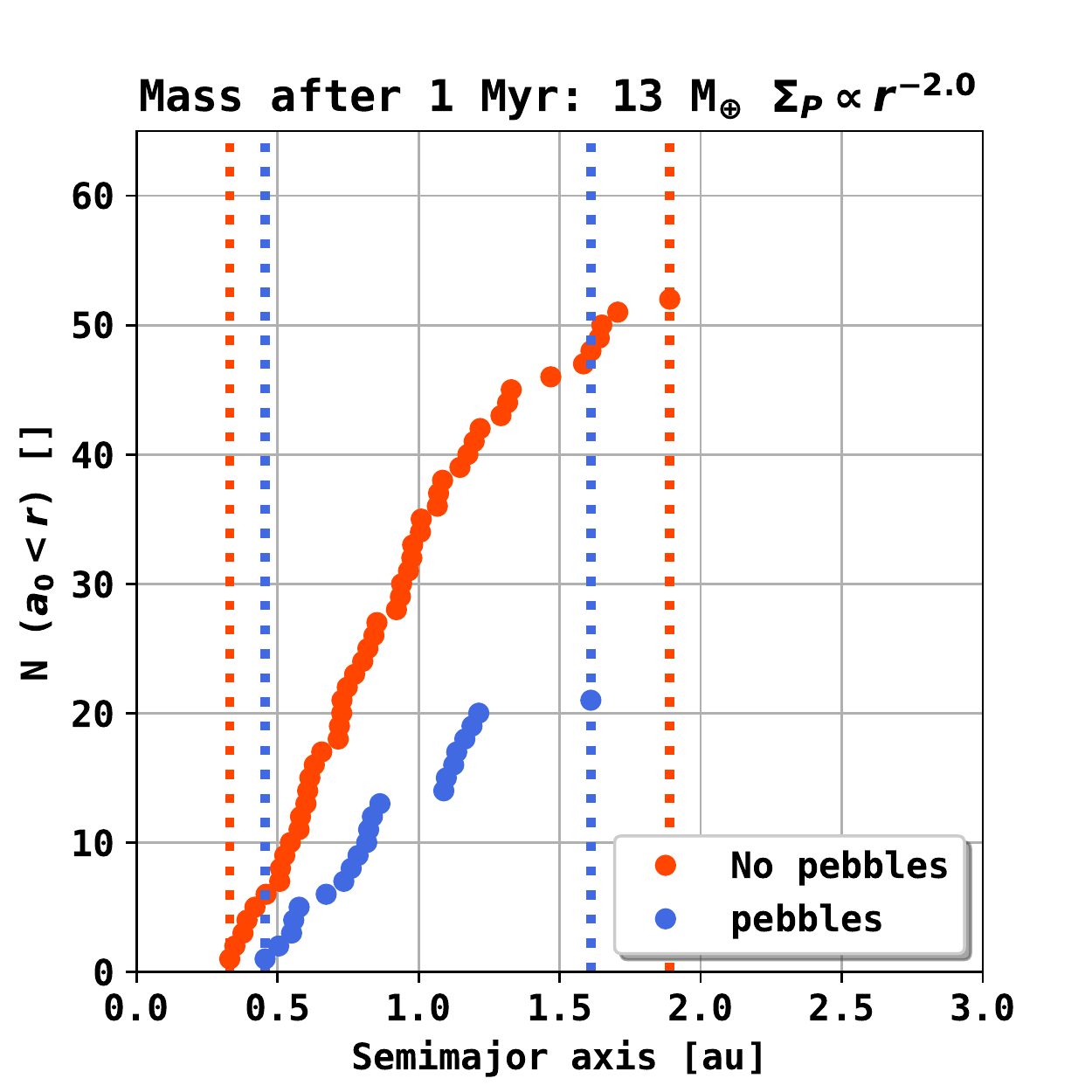}
\end{minipage}%
\\
\begin{minipage}{.33\textwidth}
  \centering
  \includegraphics[width=1.0\linewidth]{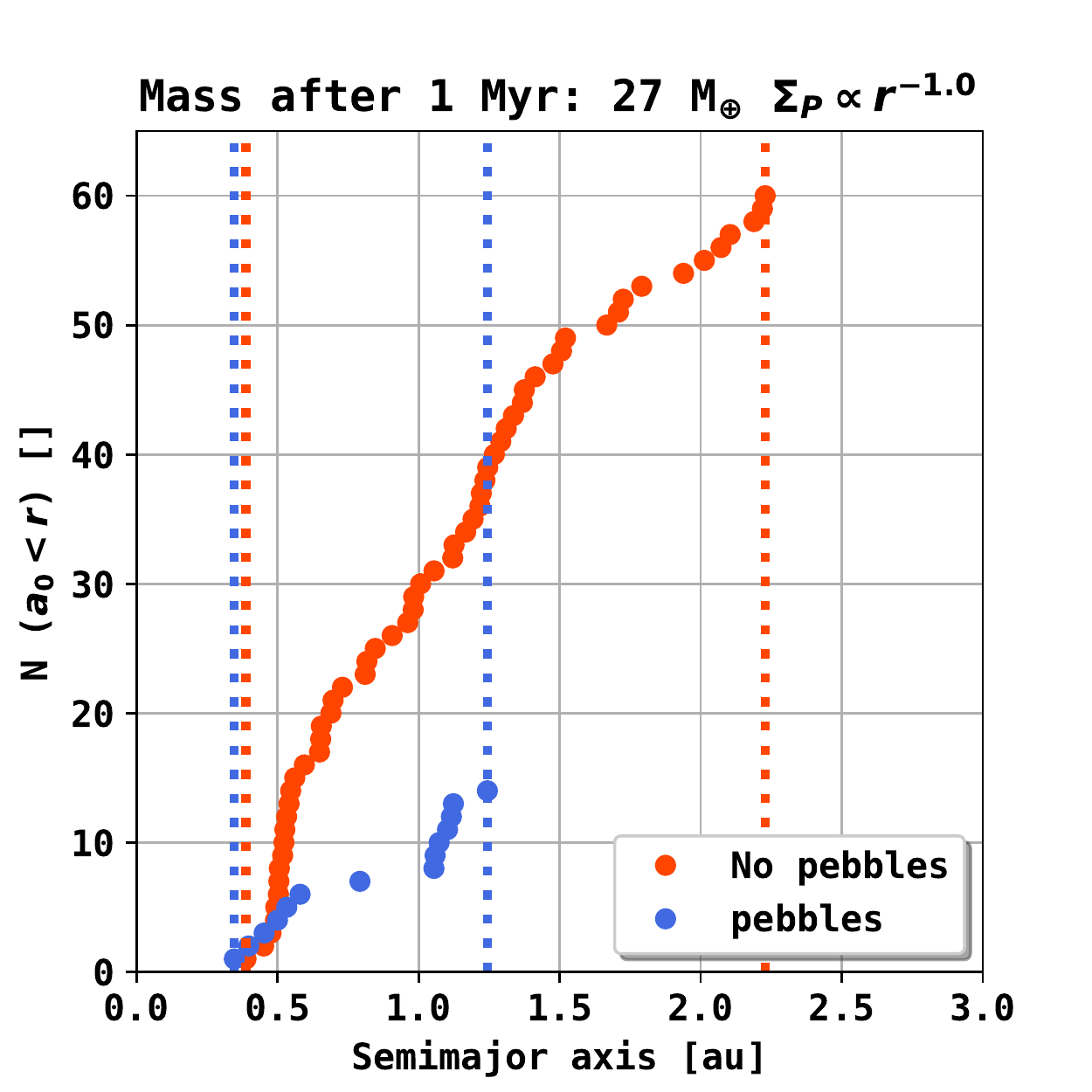}
\end{minipage}
\begin{minipage}{.33\textwidth}
  \includegraphics[width=1.0\linewidth]{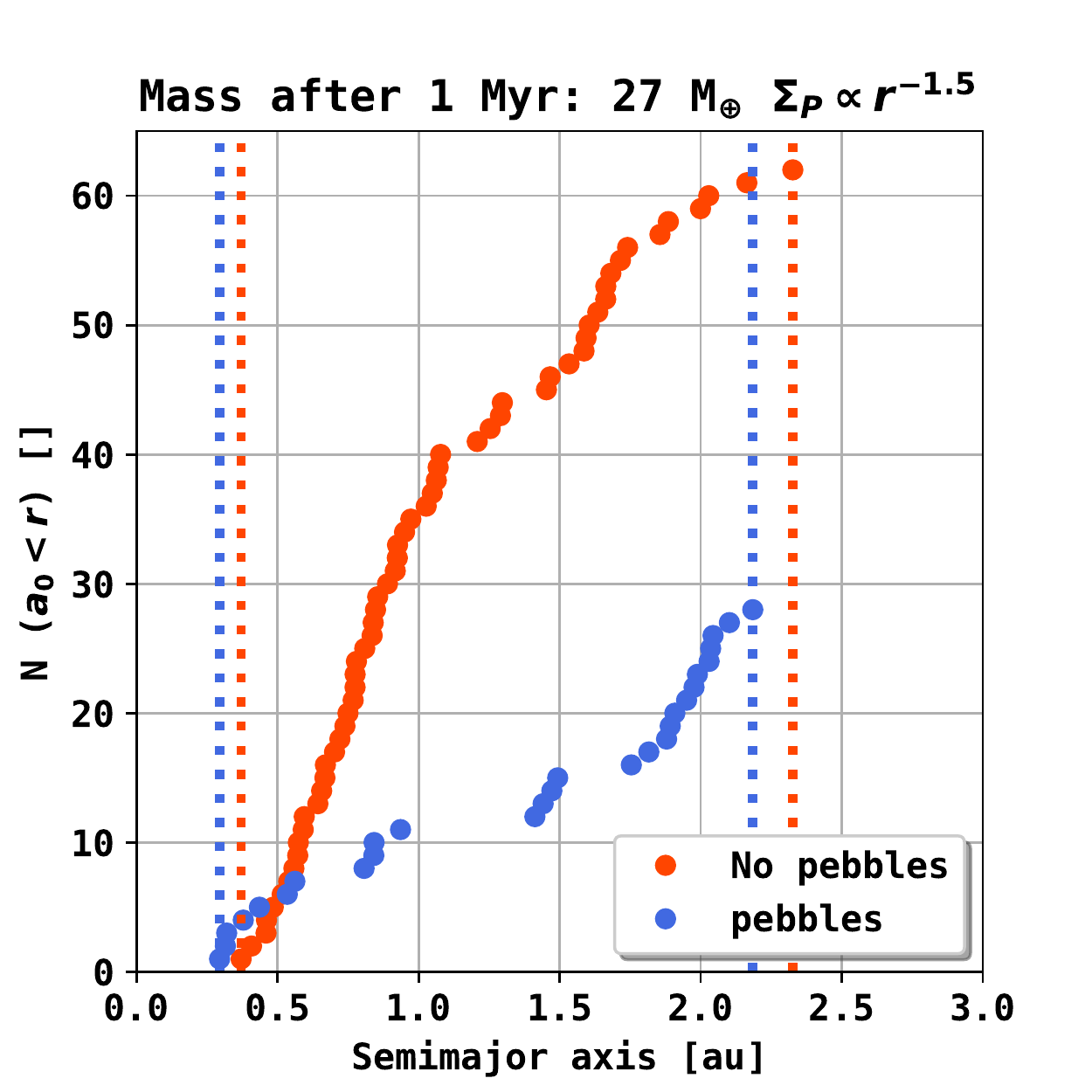}
\end{minipage}%
\begin{minipage}{.33\textwidth}
  \includegraphics[width=1.0\linewidth]{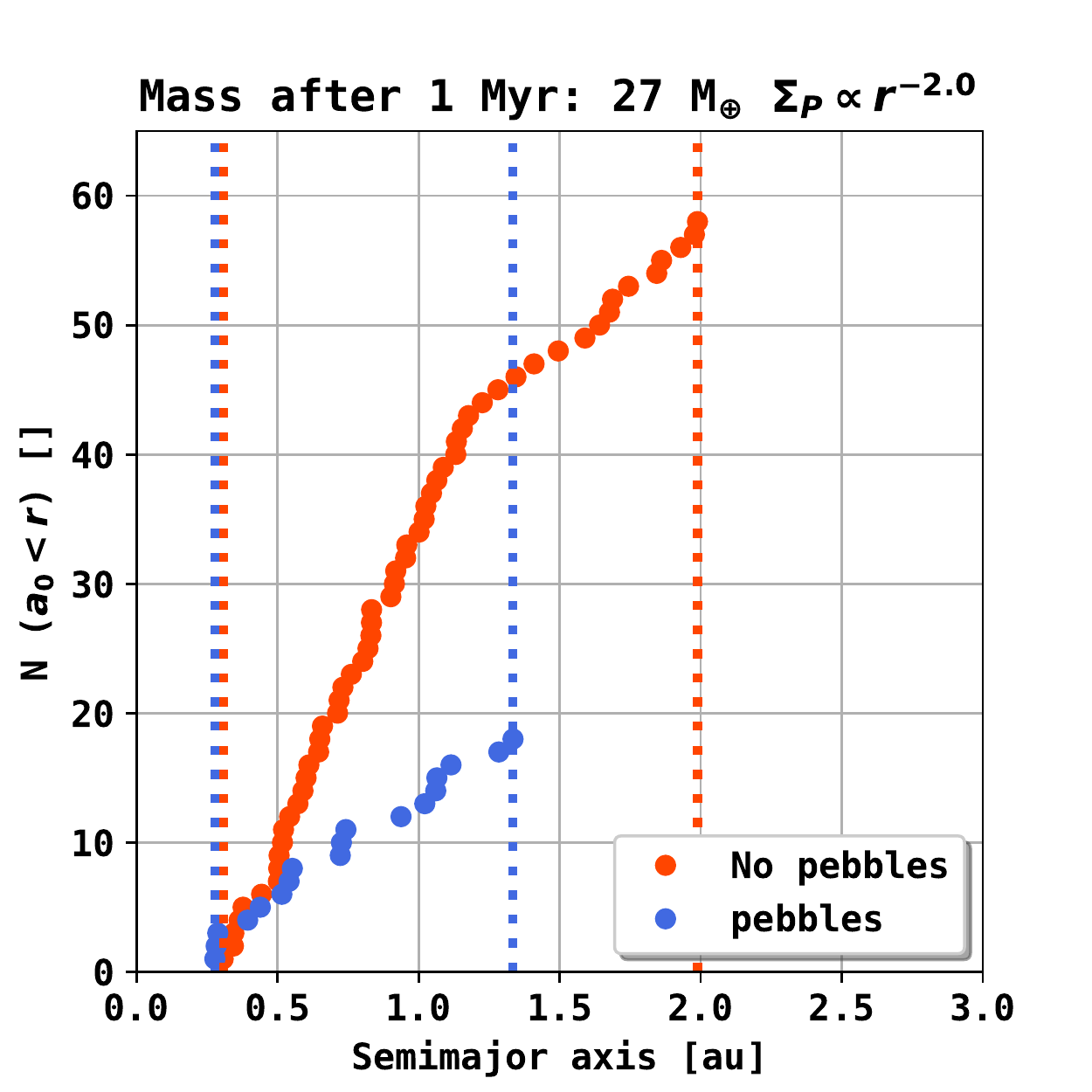}
\end{minipage}%
\caption{\small Cumulative number of initial embryos after 1 Myrs for the systems from Fig. \ref{Fig:Emb_form_LIPAD_6_ME} - Fig. \ref{Fig:Emb_form_LIPAD_27_ME}. The orange dots refer to the systems in which pebble accretion is disabled, whereas the blue dots refer to the systems in which pebble accretion is enabled.
}
\label{Fig:Cumulative_number}
\end{figure*}

As already seen in Fig. \ref{Fig:Embryo_number}, the total number of active embryos in the simulation decreases strongly if pebble accretion is included. In Fig. \ref{Fig:Cumulative_number} we show the cumulative number of initial embryos for the systems from Fig. \ref{Fig:Emb_form_LIPAD_6_ME} - Fig. \ref{Fig:Emb_form_LIPAD_27_ME}. The cumulative number without pebble accretion is shown by the orange dots, the blue dots refer to the simulations including pebble accretion. We also highlight where the innermost and outermost embryos form within 1 Myrs for each setup via vertical dotted lines with corresponding colors. We find that in terms of the initial formation of embryos, the outermost embryo forms further out in the system in which pebble accretion is neglected. For the formation of the innermost embryo, pebble accretion shows no dominant effect. Since the orbital separation is still the same in terms of the embryos Hill radii, that scales linearly with the distance to the star, we find the same logarithmic distribution of cumulative embryos, but with a lower total number as in the simulations without pebble accretion.
\section{Discussion}
\label{Sec:Discussion}
\subsection{The impact of pebble accretion}
\label{Subsec:pebble_accretion_impact}
We show that an active pebble flux has major consequences on the evolution of the planetary systems within the first 1 Myr. The accretion of pebbles leads to the formation of a lower number of substantially more massive embryos within a smaller semimajor axis interval of embryo formation. The physical spacing between embryos increases due to their higher masses in the pebble accretion runs. Their orbital separation when expressed in Hill radii remains unaffected and converges to $\approx$10$\,$R$_{Hill}$ in both cases. Embryos that form at larger distances (>1.5$\,$au) well after $T_{M_{disk}> 90\%}$ remain at low masses as they fail to undergo significant pebble accretion. This behavior was already predicted in our first study that neglected the accretion of pebbles but suggested that the disk formation rate is a valid constraint for pebble accretion due to its pebble flux dependency. We find that the outer edge of embryo formation moves slightly inwards when considering the accretion of pebbles. The formation of embryos at larger heliocentric distances within the lifetime of the pebble flux does not occur. The necessary size for significant pebble accretion is not reached at larger distances within the lifetime of our pebble flux. 
\\
The formation of the first embryo occurs earlier in the inner region if pebble accretion is considered and the embryos that form first end up to have the highest masses after 1 Myr. The accretion of pebbles plays a major role once embryos have formed. Their impact on the local formation time, while noticeable, plays a subordinate role.  
\\
Generally, we can say that the accretion of pebbles strongly favors the formation of super earths in the terrestrial planet region, but it does not enhance planetary embryo formation at larger distances. 
\subsection{Consequences for the analytic embryo formation model}
\label{Subsec:toy_model}
In part I of our study we introduced an analytic model that succeeded in reproducing the results of the N-body simulations without pebble accretion. We refer to the local formation time, the spatial distribution and the total number of initial embryos. In brief summary, the formation of embryos in the analytic model is based on two Criteria. Criterion I refers to the necessary local growth time. Criterion II determines the orbital separation to other embryos. The model uses a parameterized approach to compute the local growth time scales of planetesimals based on the local planetesimal surface density evolution. Embryos are placed if the analytic growth surpassed the mass of a planetary embryo and the orbital separation to the other already existing embryos is above an input parameter. 
\\
As discussed in Sect. \ref{Subsec:pebble_accretion_impact}, the impact of pebble accretion is largely found in the mass of the embryos, not in their initial formation time. Criterion I of the embryo formation model will therefore still give the right results (even though we find slight deviations in the inner regions). 
\\
The number of embryos and their spatial distribution are determined by criterion II. Under the assumption that the already placed embryos grow respectively by pebble accretion, their Hill radii increase. The physical spacing between the embryos thus enlarges. As a consequence, the total number of embryos decreases since the semimajor axis interval of embryo formation does not increase (Criterion I). The analytic model for embryo formation from part I of our study is therefore still valid in a framework that includes pebble accretion.
\\
Implementing the analytic model into a global model for planet formation that includes planetesimal formation and pebble accretion will be subject to future studies.  
\section{Summary and Outlook}
\label{Sec:Summary}
We study the impact of pebble accretion and planetesimal formation on the formation of planetary embryos in the terrestrial planet region. For this purpose we connected a one dimensional model for pebble flux regulated planetesimal formation and solid evolution with the N body code LIPAD. Thus we studied the growth and fragmentation of planetesimals with an initial size of 100$\,$km in diameter within the first million years of a viscously evolving circumstellar disk. In this paper we compare 18 different N-body simulations in which we vary the total mass in planetesimals, the radial pebble flux and the planetesimal surface density profile. 
Building on the efforts of our previous study \citep{voelkel2020embI} we include a radial pebble flux and the accretion of pebbles during the formation of planetary embryos. 
The main impacts on embryo formation by pebble accretion in the terrestrial planet region can be summarized as follows:
\begin{itemize}
\item Pebble accretion is highly beneficial for the formation of super earths.
\\
\item When compared with planetesimal accretion alone, the total number of embryos decreases strongly if pebble accretion is considered, while the individual embryos grow significantly more massive.
\\
\item Embryos that form early in the inner regions of the disk grow rapidly by pebble accretion, whereas the outer embryos that form later fail to do so.
\\
\item The outer edge of planetary embryo formation is not increased if pebble accretion is included. Our work indicates that it is not possible to form planetary embryos at larger distances (>2au) within the lifetime of a radial pebble flux for our assumptions.
\end{itemize}
Our findings from the first part of our study are still valid, the formation of planetary embryos occurs first in the innermost regions and then proceeds to larger distances. The number of embryos is given as the number of orbital distances within their possible formation zone. Since embryos grow more massive when pebble accretion is included, we find that the number of embryos decreases. The area in which they form however is not increased by pebble accretion since pebble accretion only becomes an effective growth mechanism for much larger sizes than 100$\,$km. By the time the outer objects have grown to larger sizes by planetesimal collisions, the pebble flux has largely vanished.
 Even though we can see that the first embryos form earlier in the inner parts of the disk for the simulations in which pebbles are accreted, this trend does not continue to larger distances. The conundrum of distant embryo formation within the lifetime of a radial pebble flux as found in part I of our study \citep{voelkel2020embI} remains. A possible solution to this issue could be in the form of locally enhanced substructures in the planetesimal surface density profile at larger distances or the formation of planetesimals that initially form large enough for pebble accretion. Future work will include the formation of planetary embryos in distant local substructures, like in pressure bumps and around the water iceline \citep{Drazkowska2017}. 
\section*{Acknowledgements}
\bibliography{Template.bib}
\end{document}